\documentclass[]{aastex7}
\usepackage{lineno}
\linenumbers

\newcommand{\Lya}     {Ly$\alpha$}    
\newcommand {\HI}        {\ion{H}{1}}     

\received{December 19, 2024}
\revised{March 10, 2025}
\accepted{March 11, 2025}
\submitjournal{AJ}

\shorttitle{The Ly$\alpha$ Sky as Observed by New Horizons}
\shortauthors{Gladstone et~al.}
\watermark{draft}

\begin{document}
\title{The Ly$\alpha$ Sky as Observed by New Horizons at 57 AU}

\correspondingauthor{G. Randall Gladstone}
\email{grgladstone@gmail.com}

\author{G. Randall Gladstone}
\affiliation{Southwest Research Institute, 6220 Culebra Road, San Antonio, TX 78238, USA}
\affiliation{University of Texas at San Antonio, One UTSA Circle, San Antonio, TX 78249, USA}
\email{grgladstone@gmail.com}

\author{J. Michael Shull}
\affiliation{Department of Astrophysical \& Planetary Sciences, University of Colorado, Boulder CO 80309, USA}
\affiliation{Department of Physics \& Astronomy, University of North Carolina, Chapel Hill, NC 27599, USA}
\email{Michael.Shull@Colorado.EDU}

\author{Wayne R. Pryor}
\affiliation{Central Arizona College, 8470 North Overfield Road, Coolidge, AZ 85128, USA}
\email{Wayne.Pryor@centralaz.edu}

\author{Jonathan Slavin}
\affiliation{Harvard-Smithsonian Center for Astrophysics, 60 Garden Street, Cambridge, MA 02138, USA}
\email{jslavin@cfa.harvard.edu}

\author{Joshua A. Kammer}
\affiliation{Southwest Research Institute, 6220 Culebra Road, San Antonio, TX 78238, USA}
\email{jkammer@swri.edu}

\author{Tracy M. Becker}
\affiliation{Southwest Research Institute, 6220 Culebra Road, San Antonio, TX 78238, USA}
\email{tbecker@swri.edu}

\author{Tod R. Lauer}
\affiliation{NSF's National Optical Infrared Astronomy Research Laboratory, P.O. Box 26732, Tucson, AZ 85726, USA}
\email{tod.lauer@noirlab.edu}

\author{Marc Postman}
\affiliation{Space Telescope Science Institute, 3700 San Martin Drive, Baltimore, MD 21218, USA}
\email{postman@stsci.edu}

\author{John R. Spencer}
\affiliation{Southwest Research Institute, 1301 Walnut Street, Boulder, CO 80302, USA}
\email{spencer@boulder.swri.edu}

\author{Joel Wm. Parker}
\affiliation{Southwest Research Institute, 1301 Walnut Street, Boulder, CO 80302, USA}
\email{joel@boulder.swri.edu}

\author{Kurt D. Retherford}
\affiliation{Southwest Research Institute, 6220 Culebra Road, San Antonio, TX 78238, USA}
\affiliation{University of Texas at San Antonio, One UTSA Circle, San Antonio, TX 78249, USA}
\email{kretherford@swri.edu}

\author{Michael A. Velez}
\affiliation{University of Texas at San Antonio, One UTSA Circle, San Antonio, TX 78249, USA}
\affiliation{Southwest Research Institute, 6220 Culebra Road, San Antonio, TX 78238, USA}
\email{michael.velez@contractor.swri.edu}

\author{Maarten H. Versteeg}
\affiliation{Southwest Research Institute, 6220 Culebra Road, San Antonio, TX 78238, USA}
\email{mversteeg@swri.edu}

\author{Michael W. Davis}
\affiliation{Southwest Research Institute, 6220 Culebra Road, San Antonio, TX 78238, USA}
\email{mdavis@swri.edu}

\author{Cynthia S. Froning}
\affiliation{Southwest Research Institute, 6220 Culebra Road, San Antonio, TX 78238, USA}
\email{cfroning@swri.edu}

\author{Camden D. Ertley}
\affiliation{Southwest Research Institute, 6220 Culebra Road, San Antonio, TX 78238, USA}
\email{certley@swri.edu}

\author{Nathaniel Cunningham}
\affiliation{Physics Department, Nebraska Wesleyan University, Lincoln, NE 68504, USA}
\email{nathaniel@boulder.swri.edu}

\author{Jayant Murthy}
\affiliation{Indian Institute of Astrophysics, Bengaluru, India}
\email{jmurthy@yahoo.com}

\author{Richard C. Henry}
\affiliation{Henry A. Rowland Department of Physics \& Astronomy, The Johns Hopkins University, Baltimore, MD 21218, USA}
\email{henry@jhu.edu}

\author{Seth Redfield}
\affiliation{Astronomy Department, Wesleyan University, Middletown, CT 06459, USA}
\email{sredfield@wesleyan.edu}

\author{Carey M. Lisse}
\affiliation{The Johns Hopkins University Applied Physics Laboratory, 11100 Johns Hopkins Road, Laurel, MD 20723, USA}
\email{carey.lisse@jhuapl.edu}

\author{Kelsi N. Singer}
\affiliation{Southwest Research Institute, 1301 Walnut Street, Boulder, CO 80302, USA}
\email{ksinger@swri.edu}

\author{Anne J. Verbiscer}
\affiliation{University of Virginia, Charlottesville, VA 22904, USA}
\email{av4n@virginia.edu}

\author{Pontus C. Brandt}
\affiliation{The Johns Hopkins University Applied Physics Laboratory, 11100 Johns Hopkins Road, Laurel, MD 20723, USA}
\email{Pontus.Brandt@jhuapl.edu}

\author{S. Alan Stern}
\affiliation{Southwest Research Institute, 1301 Walnut Street, Boulder, CO 80302, USA}
\email{alan@boulder.swri.edu}




\begin{abstract}

During September~2023 the Alice ultraviolet spectrograph on the New Horizons (NH)
spacecraft was used to map diffuse Lyman alpha (Ly$\alpha$) emission over most of the sky,
at a range of $\sim56.9$~AU from the Sun.
At that distance, models predict that the interplanetary medium Ly$\alpha$ emissions result
from comparable amounts of resonant backscattering of the solar Ly$\alpha$ line by interstellar
hydrogen atoms (\HI) passing through the solar system, in addition to an approximately isotropic background
of $\sim50\pm20$~R from the Local InterStellar Medium (LISM).
The NH observations show no strong correlations with nearby cloud structures of the LISM or
with expected structures of the heliosphere, such as a hydrogen wall associated with the heliopause.
To explain the relatively bright and uniform Ly$\alpha$ of the LISM we propose that hot, young stars
within the Local Hot Bubble (LHB) shine on its interior walls, photoionizing \HI\ atoms there.
Recombination of these ions can account for the observed $\sim50$~R Ly$\alpha$ background,
after amplification of the diffuse Ly$\alpha$ by resonant scattering, although
sophisticated (i.e., 3-D) radiative transfer models should be used to confirm this conjecture.
Future observations of the diffuse Ly$\alpha$, with instruments capable of resolving the line profile,
could provide a new window on \HI\ populations in the LISM and heliosphere.
The NH Alice all-sky Ly$\alpha$ observations presented here may be repeated at some point in
the future, if resources allow, and the two maps could be combined to provide a significant
increase in angular resolution.

\end{abstract}



\section{Introduction}

Neutral hydrogen atoms (\HI) are ubiquitous in the universe, with a primary allowed electronic transition from
$nl=2p$ to $nl=1s$ emitting photons at a wavelength of 121.567~nm (i.e., Lyman~alpha, Ly$\alpha$) which are seen everywhere.
In fact, a sizable fraction of the photon energy in our galaxy (Dijkstra~2019) is thought to be carried by Ly$\alpha$ photons.
In the local interstellar medium (LISM) these Ly$\alpha$ photons are a million times more likely to undergo resonance
scattering than they are to be absorbed by dust, so that very diffuse emissions of Ly$\alpha$ are to be expected.
The local Galactic and solar system backgrounds of Ly$\alpha$ emission have been of scientific interest
since the space age began (e.g., M{\"u}nch~1962; Adams~1971; Fahr~1974).
For many years it has been understood that the Sun resides near a set of low-density clouds of neutral H atoms
(the ``local fluff'' within 10~pc) which are located inside a much larger, low-density cavity (the ``local hot bubble'',
LHB) (Lallement et~al.~1986; Cox \& Reynolds~1987; Welsh et~al.~1991).
A recent survey of molecular clouds and star-forming regions (Zucker et~al.~2022) suggested that
many of these regions lie along the boundary of the LHB.
In the inner region of the solar system, any Galactic background at Ly$\alpha$ is dominated by resonantly backscattered
solar Ly$\alpha$ emissions, both in planetary coronas, at brightnesses up to tens of kR, where 
1~R = 1~Rayleigh = $10^6$~photons~cm$^{-2}$~s$^{-1}$~$(4\pi~{\rm sr})^{-1}$,
and in the interplanetary medium (IPM), at brightnesses of 0.5-1~kR near the Earth's orbit (e.g., Bertaux \& Blamont~1971;
Meier~1977, 1991; {\O}stgaard et~al.~2003).
Both are much brighter than the Galactic background of $30-70$~R (e.g., Gladstone et~al. 2021).

In this paper, we present the first all-sky maps of diffuse LISM Ly$\alpha$ emissions, produced using the Alice ultraviolet
spectrograph on the New Horizons (NH) spacecraft at a range from the Sun of $\sim57$~AU.
The resonantly backscattered solar Ly$\alpha$ brightness falls off nearly inversely with distance from the Sun, and
at the current distance of NH it is comparable to the brightness due to the Galactic Ly$\alpha$ background.
It is found that the component of the Ly$\alpha$ emission due to backscattered solar Ly$\alpha$ emission by 
interstellar \HI\ passing through the solar system can be effectively subtracted out using models of that component,
leaving a residual brightness map of the Galactic Ly$\alpha$ background that provides a new way to study the structure
of the LISM.
The residual Ly$\alpha$ emissions are fairly uniform and are not well correlated with known LISM structures, such as
the boundaries of local clouds, or with bow-shocks which might be expected around fast-moving stars
(Shull \& Kulkarni~2023).
The observed $50\pm20$~R of residual Ly$\alpha$ emissions likely result from H atom recombination
in the walls of the LHB, following photoionization of those atoms by nearby early-type stars within
the LHB and its walls.
While the initial source of these emissions would only result in a brightness of $\sim 1$~R if the 
Ly$\alpha$ photons were optically thin, resonant scattering amplifies the brightness within the LHB and its
walls enough to make the relatively bright and isotropic emissions observed by NH.

\begin{figure}[ht]
\centering
\includegraphics[width=6.5in]{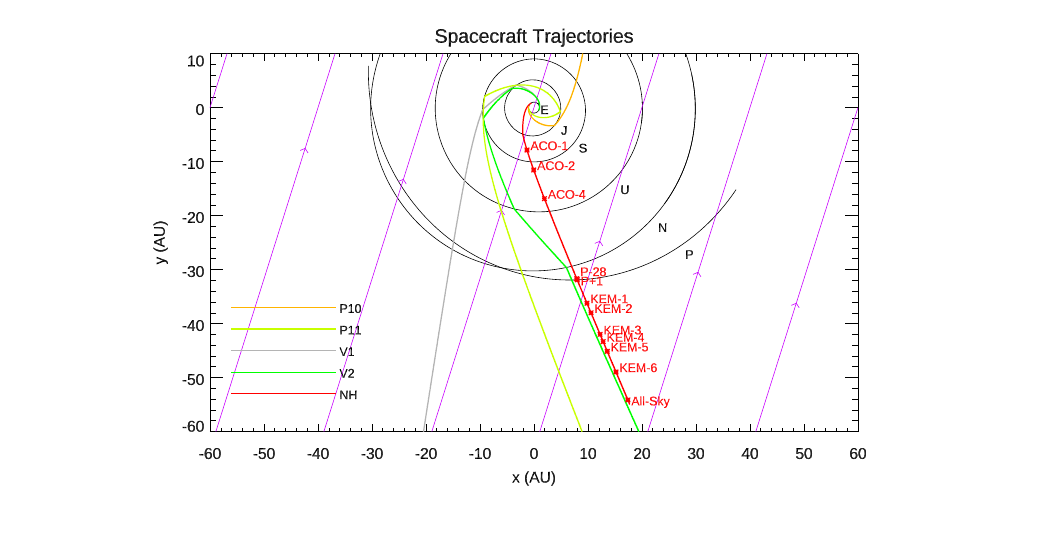}
\caption{The trajectories of the five spacecraft currently leaving the solar system:
Pioneer~10 and 11 (orange and light green, respectively); Voyager~1 and 2 (gray and green, respectively);
and New~Horizons (red) are shown projected onto the plane of the ecliptic, along with several planet
orbits (black) and the direction of the flow of interstellar hydrogen atoms (purple arrows).
The locations where great-circle scans of interplanetary medium (IPM) Ly$\alpha$ were made
with the New~Horizons Alice ultraviolet spectrograph are indicated (red), including the all-sky Ly$\alpha$ map
described here, which was executed during September 2-11, 2023 at a distance from the Sun of 56.9~AU.
\label{fig:trajectory}}
\end{figure}

\begin{figure}[ht]
\centering
\includegraphics[width=6.5in]{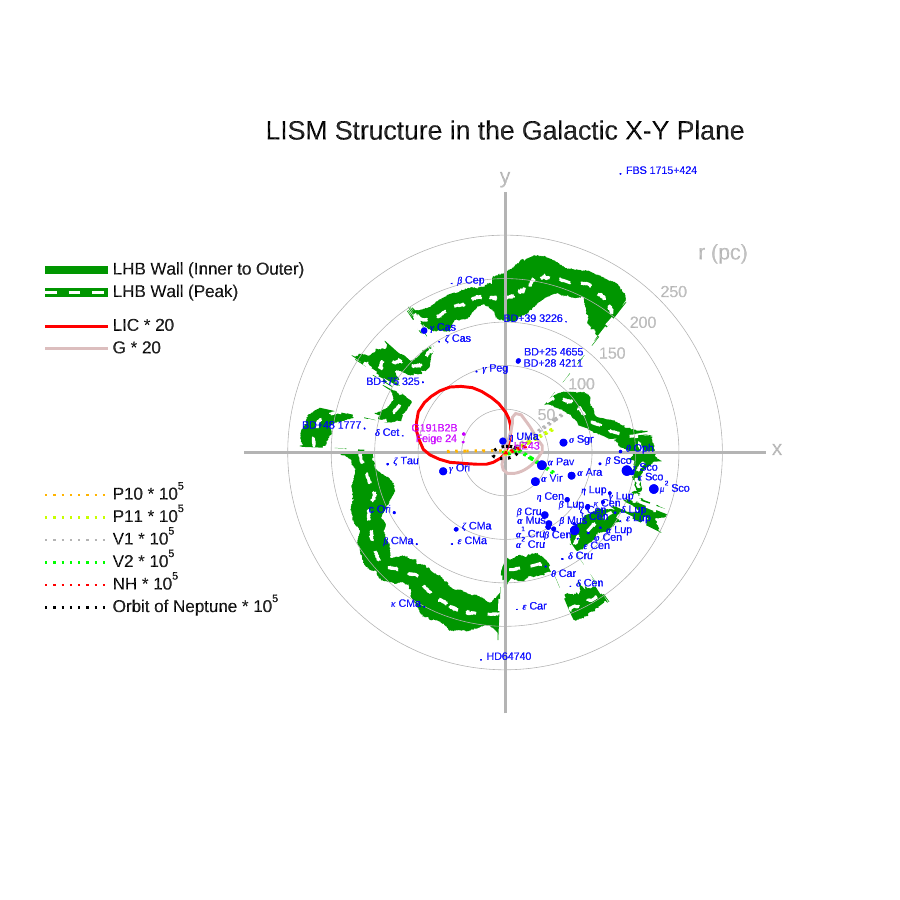}
\caption{The structure of the local interstellar medium (LISM) is shown in the $x-y$ plane of the Galaxy
(where $+x$ is directed toward ($l$,$b$) = ($0^\circ$,$0^\circ$) and $+y$ is directed toward ($90^\circ$,$0^\circ$).
The green regions show a cross section of the local hot bubble (LHB) wall as provided by O'Neill et~al. (2024), with the inner, peak,
and outer surfaces indicated.
The outlines of the LIC and G clouds (as found by Redfield \& Linsky, 2008) are shown in red and light brown, respectively,
magnified by $20 \times$.
To indicate the orientation of the solar system and for comparison with Fig.~\ref{fig:trajectory}, the trajectories of the five
escaping spacecraft and the orbit of Neptune are projected onto the $x-y$ plane, magnified by $100,000 \times$.
The locations of 54 UV-bright stars are projected onto the $x-y$ plane, with blue for O and B stars inside
the outerwall of the LHB and purple for three nearby EUV-bright white dwarfs (G191B2B, Feige 24, and HZ43).
\label{fig:lism_xy}}
\end{figure}

\begin{figure}[ht]
\centering
\includegraphics[width=6.5in]{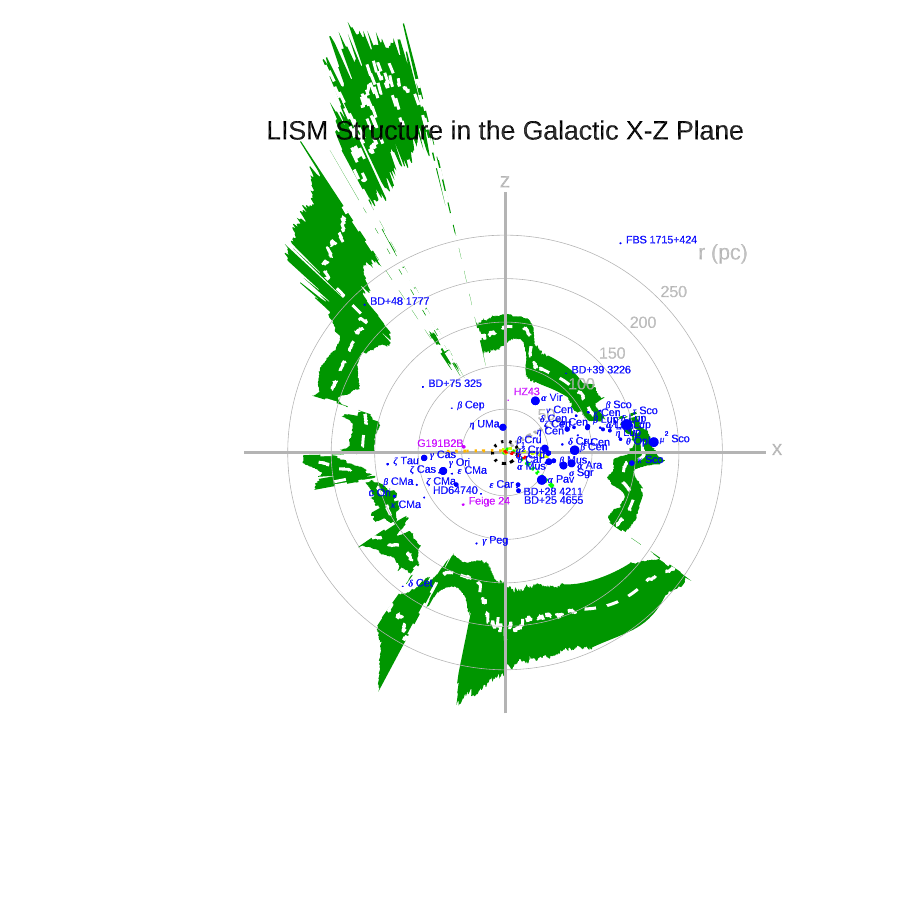}
\caption{The structure of the local interstellar medium (LISM) is shown in the $x-z$ plane of the Galaxy
(where $+x$ is directed toward ($l$,$b$) = ($0^\circ$,$0^\circ$) and $+z$ is directed toward ($0^\circ$,$90^\circ$).
The rest of the figure is as in Fig.~\ref{fig:lism_xy}, except that the outlines of the LIC and G clouds are not shown.
\label{fig:lism_xz}}
\end{figure}

\begin{figure}[ht]
\centering
\includegraphics[width=6.5in]{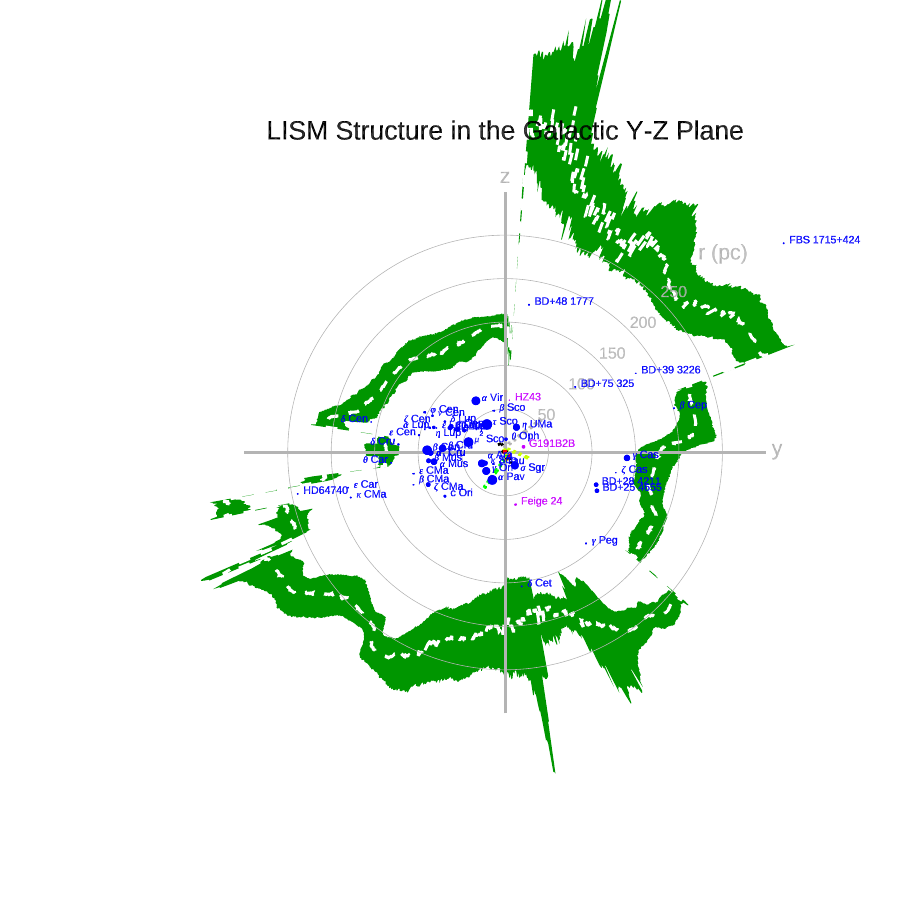}
\caption{The structure of the local interstellar medium (LISM) is shown in the $y-z$ plane of the Galaxy
(where $+y$ is directed toward ($l$,$b$) = ($90^\circ$,$0^\circ$) and $+z$ is directed toward ($0^\circ$,$90^\circ$).
The rest of the figure is as in Fig.~\ref{fig:lism_xy}, except that the outlines of the LIC and G clouds are not shown.
\label{fig:lism_yz}}
\end{figure}

\section{New Horizons Alice Data}

The distant location of the New Horizons spacecraft from the Sun provides a useful platform for observing the
diffuse Ly$\alpha$ background of the heliosphere and LISM.
Figure~\ref{fig:trajectory} shows the NH, Voyager, and Pioneer spacecraft trajectories projected onto the plane of the ecliptic,
with the flow of interstellar \HI\ through the solar system indicated for comparison.
While more sparse NH observations have been made and discussed earlier (Qu{\'e}merais et~al.~2013;
Gladstone et~al.~2013, 2015, 2018, 2021), the current results use a much more extensive, nearly all-sky
data set which was obtained during September 2-11, 2023 at a range from the Sun of 56.9~AU.

For context, Figures~\ref{fig:lism_xy}, \ref{fig:lism_xz}, and \ref{fig:lism_yz}, show some of the important
features of the LISM, e.g., some of the elements of Fig.~\ref{fig:trajectory}, walls of the LHB as recently
determined (O'Neill et~al.~2024) and nearby stars that are bright at far ultraviolet
(FUV, $91 < \lambda {\rm(nm)} < 200$) and extreme ultraviolet (EUV, $\lambda {\rm(nm)} < 91$) wavelengths (cf. Table~\ref{tab:lhbstars}).
Figure~\ref{fig:lism_xy} also indicates the orientation of the LIC and G clouds in which the Sun is embedded. 

The Alice ultraviolet spectrograph on the New Horizons spacecraft uses a 4-cm aperture off-axis telescope to feed a
Rowland-circle spectrograph, with a double-delay-line (DDL, a particular anode design, cf., Vallerga \& McPhate, 2000)
curved microchannel plate (MCP) detector at the focal plane and electronics and mechanisms in a single unit (Stern et~al. 2008).
The entrance slit has a $2^\circ \times 2^\circ$~`Box' and a $0.1^\circ \times 4^\circ$~`Slot'.
The box portion was required by the Alice design in order to assure that the radio and solar occultations at Pluto
(which overlapped in time) could be observed together.
The Alice spectral bandpass is 52--187~nm with a filled-slit spectral resolution of 0.9~nm in the $0.1^\circ$-wide slot.
The $2^\circ$-box is wide enough for off-axis diffuse Ly$\alpha$ emissions to land on KBr and CsI photocathodes which
coat the MCP on either side of a bare region of the detector where on-axis Ly$\alpha$ photons fall, giving
the spectrograph a high sensitivity to diffuse Ly$\alpha$ emissions (Gladstone et~al.~2015).
This sensitivity was initially estimated at 5.5~counts s$^{-1}$ R$^{-1}$ (Gladstone et~al.~2015), and then lowered to
$4.92\pm0.09$~counts s$^{-1}$ R$^{-1}$ (about $11\%$ less than our initial estimate) based on a more careful study 
(Gladstone et~al.~2021).
However, this previous study was not careful enough, as it was discovered in this work that the wrong background count 
rate was used in the earlier studies.
The Alice instrument produces a digital count rate for science observations (using analog counts from the MCP detector
which have been further processed by the detector electronics) for science observations, and an analog count rate for housekeeping
(using unfiltered analog counts from the MCP detector).
It is this analog count rate that is used in the great-circle Ly$\alpha$ observations, in order to save on downlinked data
volume, a precious resource at the large distance from Earth (and corresponding low data transmission rate) of NH.
Since the diffuse Ly$\alpha$ emissions dominate the count rate when no bright stars are in the Alice field of view,
we operate the spectrograph as a photometer, and downlink only the total analog count rate observed, with no spectral information.
In previous great-circle observations, the data were sent to the ground as housekeeping data, 
but for the much larger all-sky map discussed here, new flight software was developed to use science frames to hold
the analog count rate data, providing a large savings in data volume.
This new software also allows sampling at a higher rate than the maximum 1-Hz rate at which housekeeping data can be taken.
For the all-sky map, the sampling rate was set at 10~Hz.
Each sample uses 8~bits, so the maximum count rate is 2550~counts~s$^{-1}$. 
Adjusting for the analog instrument dead time of 4~$\mu$s, the true maximum count rate for the Alice all-sky map is
2576~counts~s$^{-1}$.

Using the proper background, the current estimate of the Alice sensitivity is $3.67\pm0.02$~counts s$^{-1}$ R$^{-1}$, $\sim25\%$
less than used in Gladstone et~al. (2021), and the derivation of this value is described in Appendix~A.
This sensitivity is still $\sim350 \times$ larger than the corresponding sensitivity of the Voyager~UVS
spectrometers (due to the large solid angle of the Alice box and the high yields of the KBr and CsI photocathodes).
 
The Alice all-sky map was made by expanding on the six-great-circle concept described in previous studies
(Gladstone et~al.~2015, 2021).
The sky was scanned in five contiguous $30^\circ$ segments comprising 15 great circles, with each great circle
spaced $2^\circ$ from the previous one.
The scan rate in the direction perpendicular to the Alice slot was set at $0.05^\circ$ s$^{-1}$, so that each of 75 great circle
observation lasted 2~hours and the entire observation took 150~hours.
The $30^\circ$ segment centered on the Sun (and anti-Sun) was not observed (this would have been segment~3), to
protect Alice and co-boresighted instruments from the brightness of the Sun itself.

Some details of the all-sky map segments are provided in Table~\ref{tab:obs}, which includes start and end times
in New Horizons Mission Elapsed Time (MET) and Coordinated Universal Time (UTC), scan duration,
and the R.A. \& Dec. of the New Horizons spacecraft Z-axis (the great circle axis) at the start of the segment.

\begin{deluxetable*}{cccccccc}
\tablecaption{Circumstances of New~Horizons Alice All-Sky Ly$\alpha$ Observations\label{tab:obs}}
\tablewidth{700pt}
\tabletypesize{\scriptsize}
\tablehead{
\colhead{Segment$^a$} & \colhead{Start MET} & \colhead{End MET} & \colhead{Start UTC} & \colhead{End UTC} &
\colhead{Duration} & \colhead{Z-axis R.A.$^b$} & \colhead{Z-axis Dec.$^b$} \\
 & (s) & (s) & (yyyy-Mon-dd hh:mm:ss) & (yyyy-Mon-dd hh:mm:ss) & (hours) & ($\deg$) & ($\deg$)
} 
\startdata
6 & 555974547 & 556082974 & 2023-Sep-02 15:30:31 & 2023-Sep-03 21:37:38 & 30.11 & 125.44 & 21.64 \\
1 & 556084937 & 556193374 & 2023-Sep-03 22:10:21 & 2023-Sep-05 04:17:38 & 30.12 & 92.54 & 19.48 \\
2 & 556194737 & 556303174 & 2023-Sep-05 04:40:21 & 2023-Sep-06 10:47:38 & 30.12 & 36.57 & 1.32 \\
4 & 556543636 & 556652074 & 2023-Sep-09 05:35:20 & 2023-Sep-10 11:42:38 & 30.12 & 6.56 & -10.01 \\
5 & 556653437 & 556761874 & 2023-Sep-10 12:05:21 & 2023-Sep-11 18:12:38 & 30.12 & 336.19 & -17.60 \\
\enddata
\noindent $^a$The unobserved segment~3 is the segment containing the Sun. \\
\noindent $^b$New Horizons spacecraft Z-axis direction at the start of each segment. \\
\end{deluxetable*}

\begin{figure}[ht]
\centering
\includegraphics[width=6.5in]{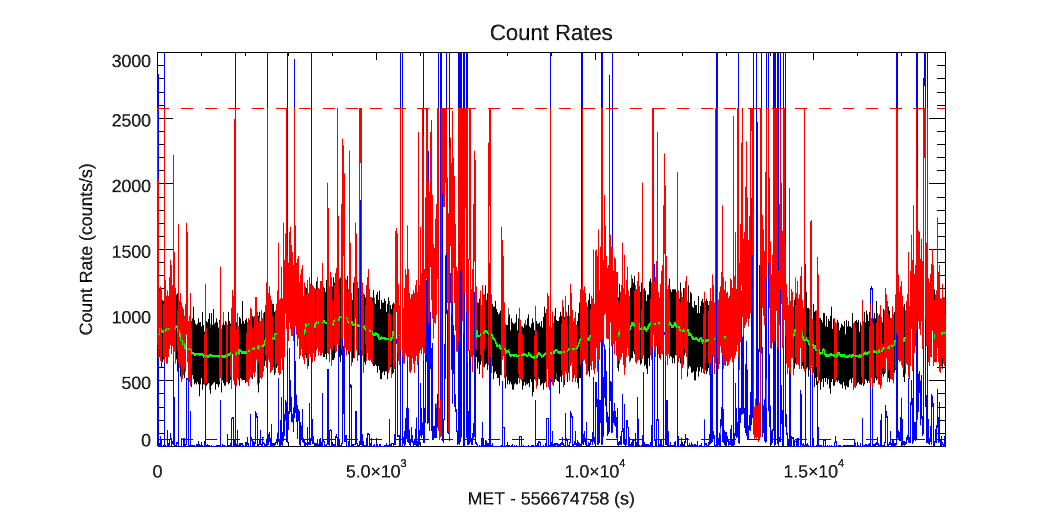}
\caption{A 5-hour excerpt from the $\sim150$~hours of data is used to illustrate how
the data were processed prior to making the all-sky map.
The raw 10-Hz count rate samples are shown in black.
The blue line shows the expected count rate due to stars passing through the Alice slit, using the
Velez et~al. (2024) star catalog.
We take any data at times where this expected stellar count rate is $>50$ counts s$^{-1}$ (shown as a horizontal
dashed blue line) as contaminated (here marked red) and remove them.
We then smooth the remaining uncontaminated data using an instrument function which is approximately a smoothed boxcar
of $\sim40$~s (i.e., $\sim400$ samples) length.  
The final count rate is shown in green, and is linearly interpolated across gaps due to star contamination.
The horizontal red dashed line indicates the maximum Alice count rate during the all-sky map of
2576~counts~s$^{-1}$.
\label{fig:decon}}
\end{figure}

At the scan speed of $0.05^\circ$ s$^{-1}$ it takes a point source $2$~s to pass through the $0.1^\circ$-width of the slot
(providing $\sim20$~samples) and $40$~s to pass through the $2^\circ$-width of the box (providing $\sim400$~samples).
In Fig.~\ref{fig:decon} we provide a close-up of just 5~hours out of the 150~hours of count rates, in which many narrow and
very narrow spikes are seen in the count rate due to stars passing through the box and slot, respectively.

While it is not easy to correct for these stars, we are able to remove most of them from the data by sampling the 
Velez et~al. (2024) star catalog in the same way that the sky was sampled by the Alice instrument and flagging times
where stars contaminated the Alice signal in a substantial way.
These contaminated data are masked and removed and the remaining uncontaminated count rates are smoothed by an instrument
function based on deep spectra of the dark sky (i.e., the NCOB observations in Table~\ref{tab:ncob} of Appendix~A).
This instrument function is basically a column-integration of the right panel of Fig.~\ref{fig:specimage}
of Appendix~A, with the x-axis converted to time (as mentioned above, the box is $\sim40$~s wide as the slit
scans a great circle on the sky).

During the all-sky scans the Alice aperture door was pre-planned to close if a particularly bright star were
passing through the slot or the box portions of the slit.
Although these times were only a small portion of the total observations ($\sim45$~minutes out of $\sim150$~hours),
they provide our best estimate of the analog dark count rate during the observations, which was determined to be 
$177$~counts s$^{-1}$ with a Gaussian distribution.
 
\begin{figure}[ht]
\centering
\includegraphics[width=6.5in]{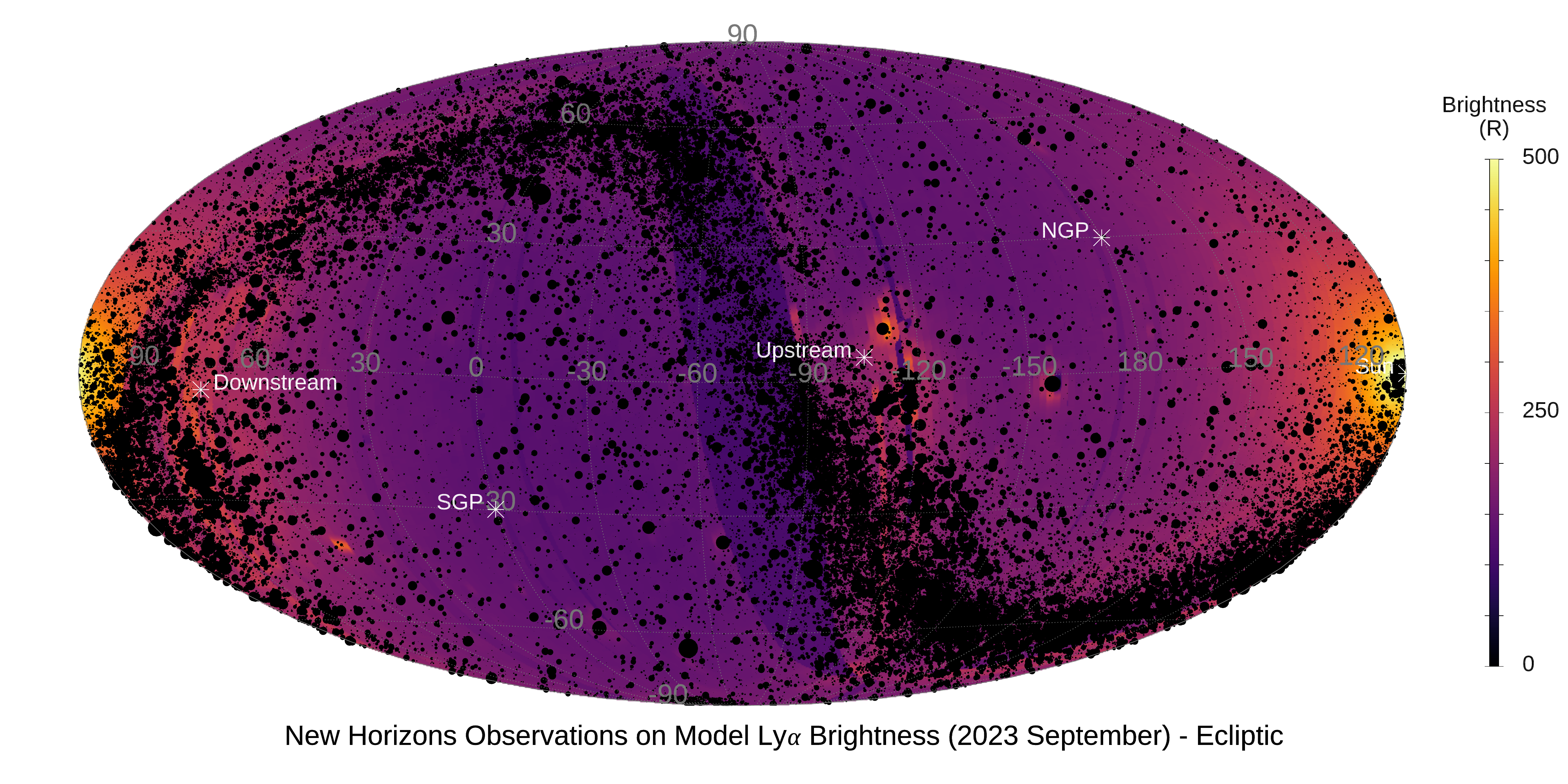}
\includegraphics[width=6.5in]{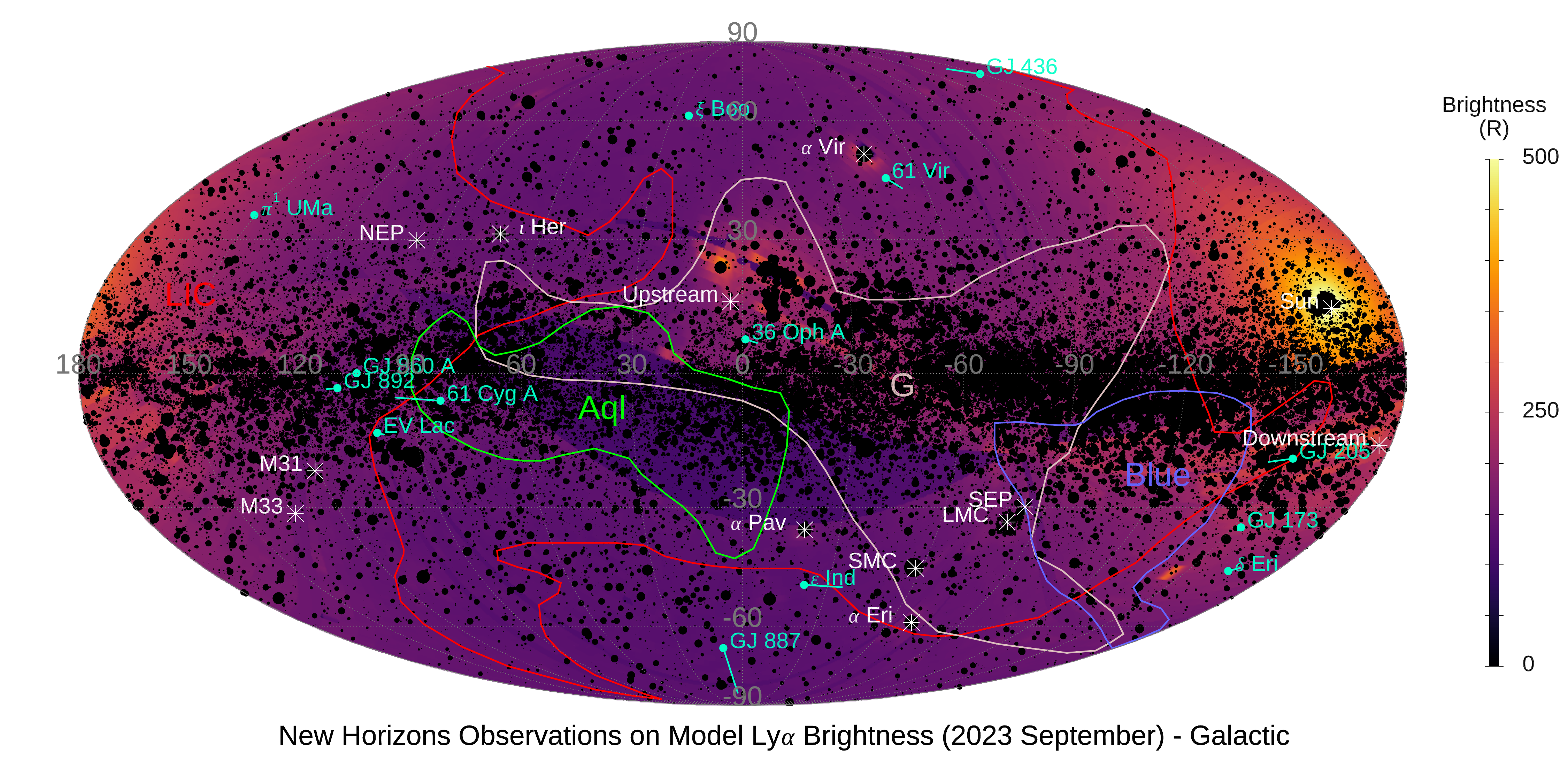}
\caption{All-sky New Horizons Alice Ly$\alpha$ maps derived from analog count rates (as processed in the
manner described in Fig.~\ref{fig:decon}) and converted to brightness using the sensitivity value of
3.67~counts~s$^{-1}$~R$^{-1}$ (1~R = $10^6$~photons cm$^{-2}$ s$^{-1}$~$(4\pi~{\rm sr})^{-1}$) that is
derived in Appendix~A.
The upper map is in ecliptic coordinates centered on the anti-Sun direction, and the lower map is
in Galactic coordinates, centered on ($l$,$b$) = ($0^\circ$,$0^\circ$). 
Both Mollweide projection maps show Alice brightnesses (except for the 30-degree swath containing the Sun, 
which shows a hot model estimate of the backscattered solar Ly$\alpha$ plus a 50-R value meant to
represent the Galactic Ly$\alpha$ signal).
The $\sim90,000$ stars in the Velez et~al. (2024) catalog are overlaid as black dots, where the size of
the dot is proportional to the logarithm of the expected Alice count rate from each star.
The map in Galactic coordinates indicates the outlines of four of the important LISM clouds
(``LIC'', in red; ``Aql'', in green; ``Blue'', in Blue; and ``G'', in tan), as described in Redfield
\& Linsky (2008).
The locations of 14 fast-moving stars from Shull \& Kulkarni (2023), listed here in Table~\ref{tab:faststars},
are indicated by light blue dots, with lines extending to their expected locations in $10^4$~years.
Both maps also include directions for the poles of the alternate projection map, the upstream and downstream
directions of the flow of interstellar \HI\ through the solar system, and the Sun.
\label{fig:lyamaps}}
\end{figure}

After the count rate data are pre-processed as described above, maps are made of the sky in both ecliptic
and Galactic coordinates, as shown in Fig.~\ref{fig:lyamaps}.
In these maps the gores (swaths of missing data on the maps) are filled with a model (descibed below) of the expected signal
from resonant backscattering of solar Ly$\alpha$ plus a $50$-R value representative of the Galactic Ly$\alpha$ background.

\section{Maps \& Models}

We build Fig.~\ref{fig:lyamaps} by first establishing a base map of the modeled sky and then overplotting the Alice observations.
The basemap can be seen in the gore spaces where we did not acquire Alice data (mostly the $30^\circ$ gore
centered on the Sun, but also the much narrower gores between segments).
The basemap model includes only an assumed constant Galactic Ly$\alpha$ background plus a
model of the resonantly backscattered solar Ly$\alpha$ brightness, which we refer to as the ``Hall'' code.
As described in Gladstone et~al. (2021), the Hall code is a hybrid of a ``hot model'' (i.e., a model of interstellar
H flow through the solar system which accounts for gravity, radiation pressure, photoionization, and H temperature, e.g.,
Thomas~1978, but doesn't account for charge exchange) of the interplanetary medium
neutral hydrogen properties, run for the specific location of NH at the time of the all-sky observations 
(e.g., Pryor et~al.~2013), and a full multiple-scattering radiative transfer model adapted from the code
described in Hall (1992) and Hall et~al. (1993).
The solar Ly$\alpha$ line profile used in the model is from Lemaire et~al. (1978).
Resonant scattering of the solar line assumes the complete frequency redistribution (CFR) approximation, with Doppler
profiles for absorption and emission.
Doppler shifts and widths in the cylindrically symmetric (about the upstream direction) coordinate system are
calculated using flow velocities and effective temperatures derived for each model volume element.
The radiative transfer equation is solved by successive orders of scattering, which is straightforward for
low optical depth problems such as scattering of solar Ly$\alpha$ in the heliosphere.

As in Gladstone et~al. (2021), the parameters for the Hall model of neutral hydrogen density, temperature, and velocity
at the outer boundary are taken to be $n_\infty = 0.12$~cm$^{-3}$ (consistent with the recent determination
of $n_\infty = 0.127\pm0.015$~cm$^{-3}$ by Swaczyna et~al.~2020), $T_\infty = 12000$~K, and $v_\infty = 20$~km s$^{-1}$
(consistent with Solar and Heliospheric Observatory (SOHO) Solar Wind ANisotropy Experiment (SWAN) hydrogen
absorption cell measurements of Costa et~al.~1999).
Likewise, the solar Ly$\alpha$ flux at the sub-New Horizons solar longitude available for scattering
from interstellar wind hydrogen, the radiation pressure parameter ($\mu$), and the total expected hydrogen atom
lifetime at 1~AU were estimated using data from the solar Ly$\alpha$ database at the University of Colorado
Laboratory for Atmospheric and Space Physics (LASP, Woods et~al. 2000).
The Hall model does not include the hydrogen wall that is expected in the outer heliosphere
by more state of the art codes (e.g., Baranov et~al.~1991; Baranov \& Malama~1993; Izmodenov et~al.~2013;
Qu{\'e}merais et~al.~2010, 2013; Katushkina et~al.~2016, 2017).
The hydrogen wall forms during charge exchange between interstellar \HI\
and interstellar protons that have been decelerated and deflected as they approach the heliopause.

\begin{figure}[ht]
\centering
\includegraphics[width=6.5in]{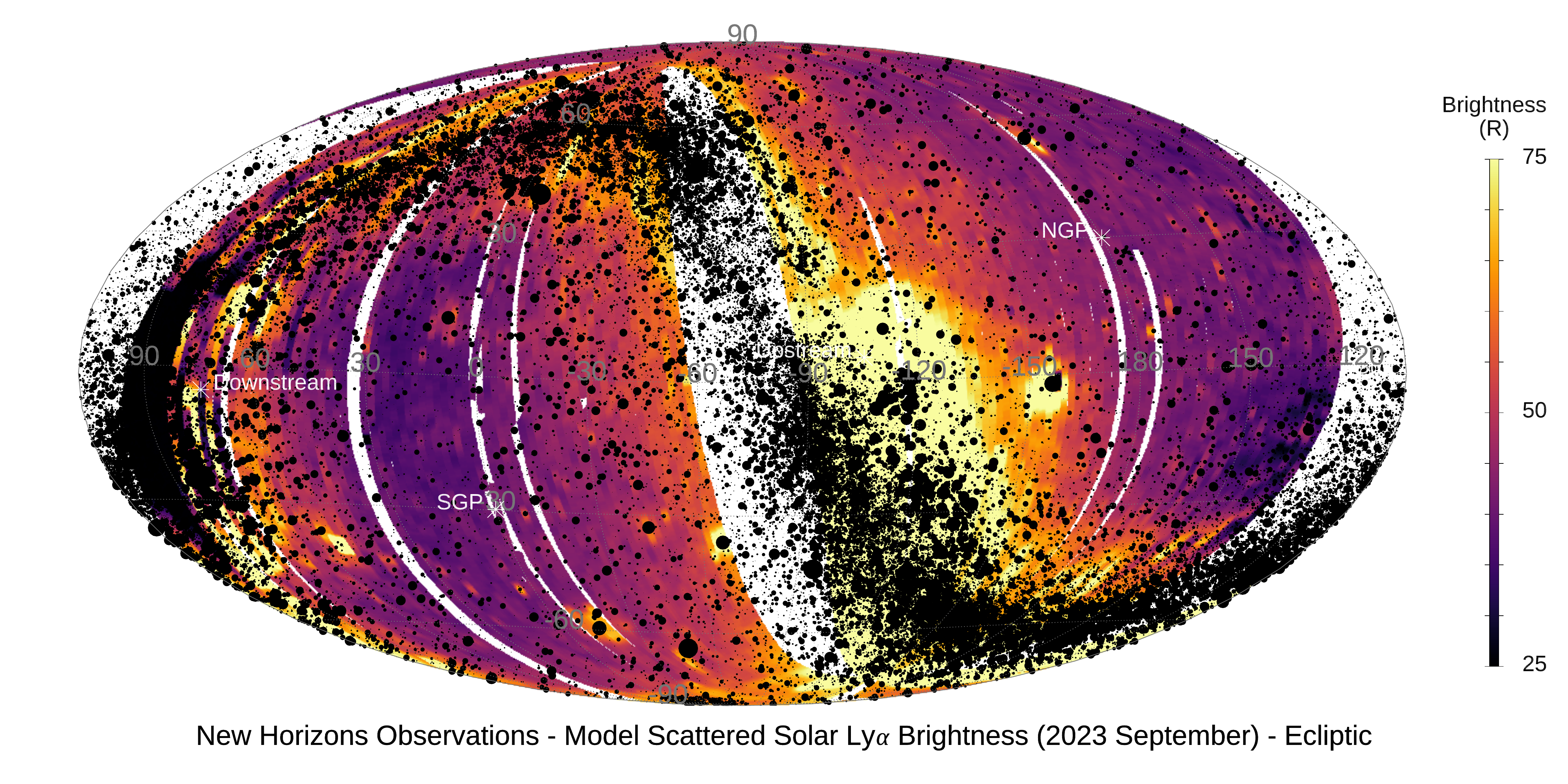}
\includegraphics[width=6.5in]{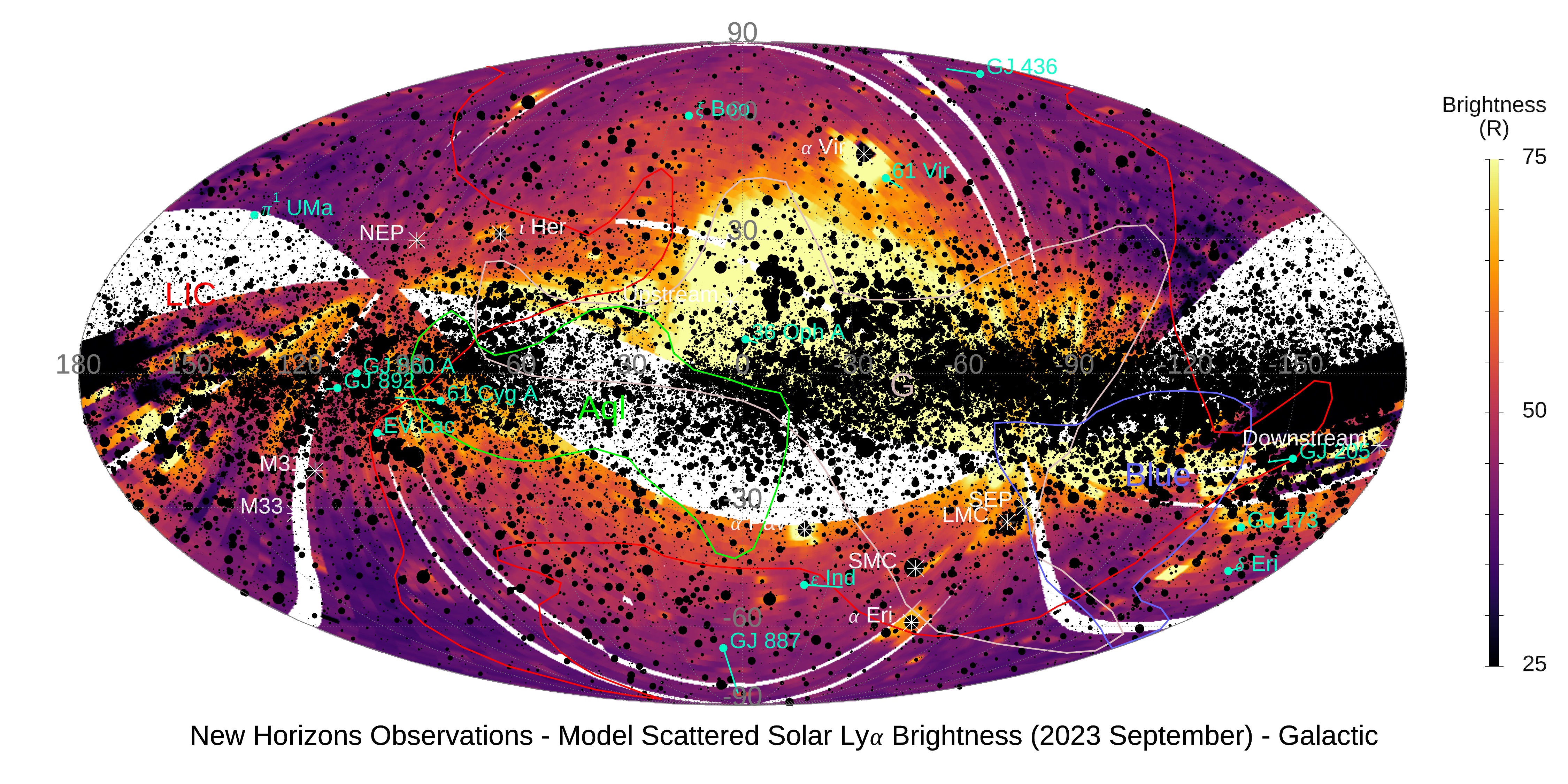}
\caption{The all-sky NH Ly$\alpha$ maps are constructed in the same way as for Fig.~\ref{fig:lyamaps}, but show the residual
brightness after subtracting a model prediction of the backscattered solar Ly$\alpha$.
Note that the range in brightness has been reduced to only $25-75$~R in these residual maps.
\label{fig:resmaps}}
\end{figure}

Since the Hall model used to estimate the backscattered solar Ly$\alpha$ and the assumption of an isotropic
Galactic background are each imperfect, we have scaled both in order to provide the best fit to the
map data of Fig.~\ref{fig:lyamaps} in the regions of the gore boundaries.
We apply a scale factor of $1.6 \times$ to the Hall model brightness map, and include an isotropic Galactic
component of $50$~R (up from the $43$~R found in Gladstone et~al.~2021).
Although this was not done in a rigorous way, the fit as shown is quite reasonable (e.g., it is hard to
see the gore boundaries, except in the anti-Sun hemisphere where the model is a few Rayleighs fainter
than the observations - the gore locations are clearly shown in Fig.~\ref{fig:resmaps}). 
Remaining differences between the model and data at the gore boundaries could reasonably be due to the Galactic
Ly$\alpha$ actually being anisotropic and/or assumptions made in the Hall model (e.g., the CFR approximation).

The maps in Fig.~\ref{fig:lyamaps} show no large-scale brightness structure that might be associated
with the presence of a hydrogen wall (e.g., Qu{\'e}merais et~al.~2010; Izmodenov et~al.~2013) in the
direction of the nose of the heliopause (i.e., centered near the location marked ``IPM Upstream'' in
the both the ecliptic and Galactic coordinates maps).
As found by Gladstone et~al. (2021), a brightness enhancement due to a hydrogen wall could be present
without detection at the level of $\sim 10$ Rayleighs (note that such a wall would be expected to be
a hemisphere-size feature).

The Galactic coordinates map in Fig.~\ref{fig:lyamaps} includes overlays of four important LISM clouds: the Local
Interstellar Cloud (LIC); the Aquilae (Aql) Cloud; the Blue Cloud; and the G Cloud; as described in
Redfield \& Linsky (2008).
None of these cloud boundaries are correlated with a noticeable change of brightness in the observed
Ly$\alpha$ brightness maps, which indicates that the cloud boundaries are gradual, at least for the
\HI\ densities they contain.

\begin{deluxetable*}{lccccccc}
\tablecaption{Nearby Fast Stars\label{tab:faststars}}
\tablewidth{700pt}
\tabletypesize{\scriptsize}
\tablehead{
\colhead{Name} &  $V_{\rm ISM}$$^a$  & \colhead{R.A.} & \colhead{Dec.} & \colhead{$l$} & 
\colhead{$b$} & \colhead{$l_{+10{\rm kyr}}$$^b$} & \colhead{$b_{+10{\rm kyr}}$$^b$} \\
  &  (km~s$^{-1}$) & ($^\circ$) & ($^\circ$) & ($^\circ$) & ($^\circ$) & ($^\circ$) & ($^\circ$)
 } 
\startdata
61 Cyg A  			& 86 & 316.725 & 38.503  & 82.135  & -5.983  & 94.518  & -5.265  \\
GJ 887    			& 85 & 346.467 & -34.147 & 9.227   & -66.310 & 4.614   & -82.560 \\
GJ 436    			& 79 & 175.546 & 26.707  & 210.542 & 74.569  & 220.986 & 76.392  \\
GJ 205    			& 70 &  82.864 & -2.323  & 205.666 & -18.818 & 212.162 & -19.573 \\
$\epsilon$ Ind    	 	& 68 & 330.840 & -55.214 & 338.317 & -48.739 & 324.524 & -49.347 \\
61 Vir    			& 51 & 199.601 & -17.689 & 311.986 & 44.708  & 307.687 & 42.089  \\
GJ 892     			& 49 & 348.321 & 57.168  & 109.899 & -3.199  & 113.089 & -3.443  \\
GJ 860 A   			& 47 & 336.752 & 57.695  & 104.575 & 0.065   & 103.326 & 0.124   \\
EV Lac    			& 45 & 341.707 & 44.334  & 100.607 & -13.069 & 98.696  & -13.493 \\
$\pi^1$ UMa   			& 43 & 129.799 & 65.021  & 150.552 & 35.704  & 150.267 & 35.621  \\
36 Oph A  			& 40 & 258.837 & -25.470 & 359.214 & 7.528   & 355.873 & 6.682   \\
GJ 173    			& 38 & 69.425  & -10.961 & 207.533 & -34.620 & 207.810 & -35.412 \\
$\delta$ Eri   			& 37 & 55.812  & -8.237  & 196.195 & -45.235 & 193.519 & -44.375 \\
$\xi$ Boo A  			& 32 & 222.847 & 19.100  & 23.086  & 61.356  & 23.128  & 61.004  \\
\enddata
\noindent $^a$ $V_{\rm ISM}$ is the ISM flow speed in the stellar rest frame (Shull \& Kulkarni~2023). \\
\noindent $^b$ Expected location of star in $10,000$~years. \\
\end{deluxetable*}

Likewise, the locations of fourteen fast-moving stars (taken from Shull \& Kulkarni~2023, and listed in Table~\ref{tab:faststars})
are overplotted as asterisks on the Galactic coordinates map in Fig.~\ref{fig:lyamaps}, along with indications
of their directions of motion.
There is no obvious indication of enhanced Ly$\alpha$ emission near these stars that might be attributed to
excitation at the bowshocks of their astrospheres (although such structures could be present at the $2-8$~R level
predicted by Shull \& Kulkarni, 2023).
Instead, the Ly$\alpha$ map is dominated by large, low-order brightness variations away from the Galactic disk,
where contamination by stars is very difficult to eliminate, and also away from the Sun.  
These regions are more extensive in the southern Galactic hemisphere than in the northern Galactic
hemisphere.
Smaller-scale features include the ``haloes'' surrounding very bright and nearby stars
(e.g., $\alpha$~Pav, $\alpha$~Eri at southern Galactic latitudes, $\alpha$~Vir, $\iota$~Her at northern
Galactic latitudes).

After scaling the backscattered solar Ly$\alpha$ to give a reasonable fit the Alice Ly$\alpha$ brightness map,
we subtract the model solar contribution (i.e., the Hall model, which is the same as the base map minus the
assumed constant 50-R background) to see the underlying structure of the Galactic Ly$\alpha$ emission.  
These residual Ly$\alpha$ maps are plotted in Fig.~\ref{fig:resmaps}.
In this figure the dynamic range of the color bar has been reduced by a factor of $10 \times$ (i.e., from $500$~R
down to $50$~R), in order to bring out details, but it is important to keep in mind that $25-75$~R is a very
small range in brightness and brings out artifacts (especially at low Galactic latitudes, where the
stellar contamination is very large).
Nevertheless, there appear to be few, if any, good correlations between the residual Ly$\alpha$ maps and
any of the structures we might expect to show up, i.e., cloud boundaries, shocks associated with fast stars, or
a hydrogen wall structure. 
The main feature of the residual all-sky Ly$\alpha$ map is its general smoothness, along with a rather high
level for the average emission brightness of $\sim50 \pm 20$~R.

\begin{figure}[ht]
\centering
\includegraphics[width=6.5in]{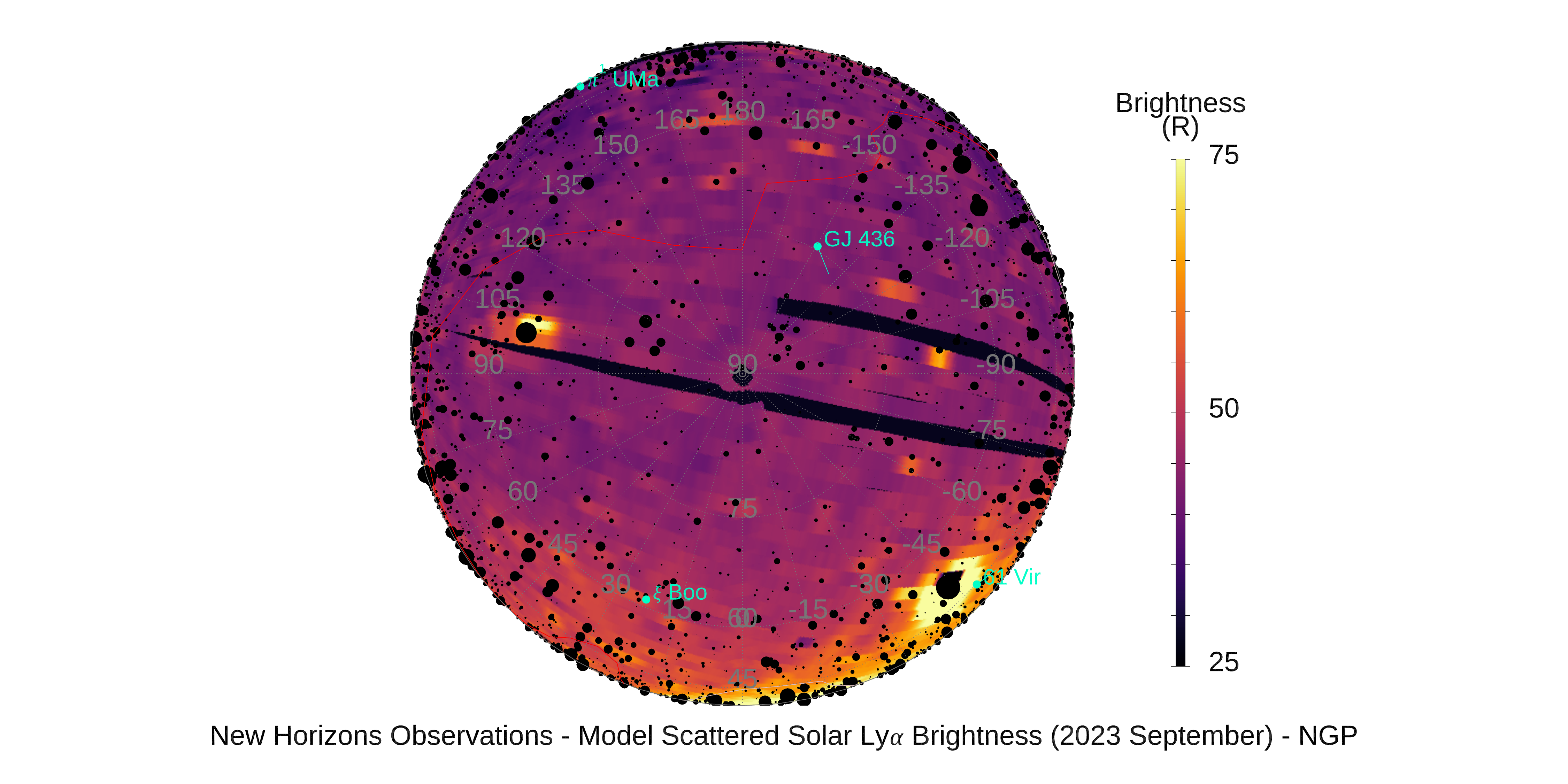}
\includegraphics[width=6.5in]{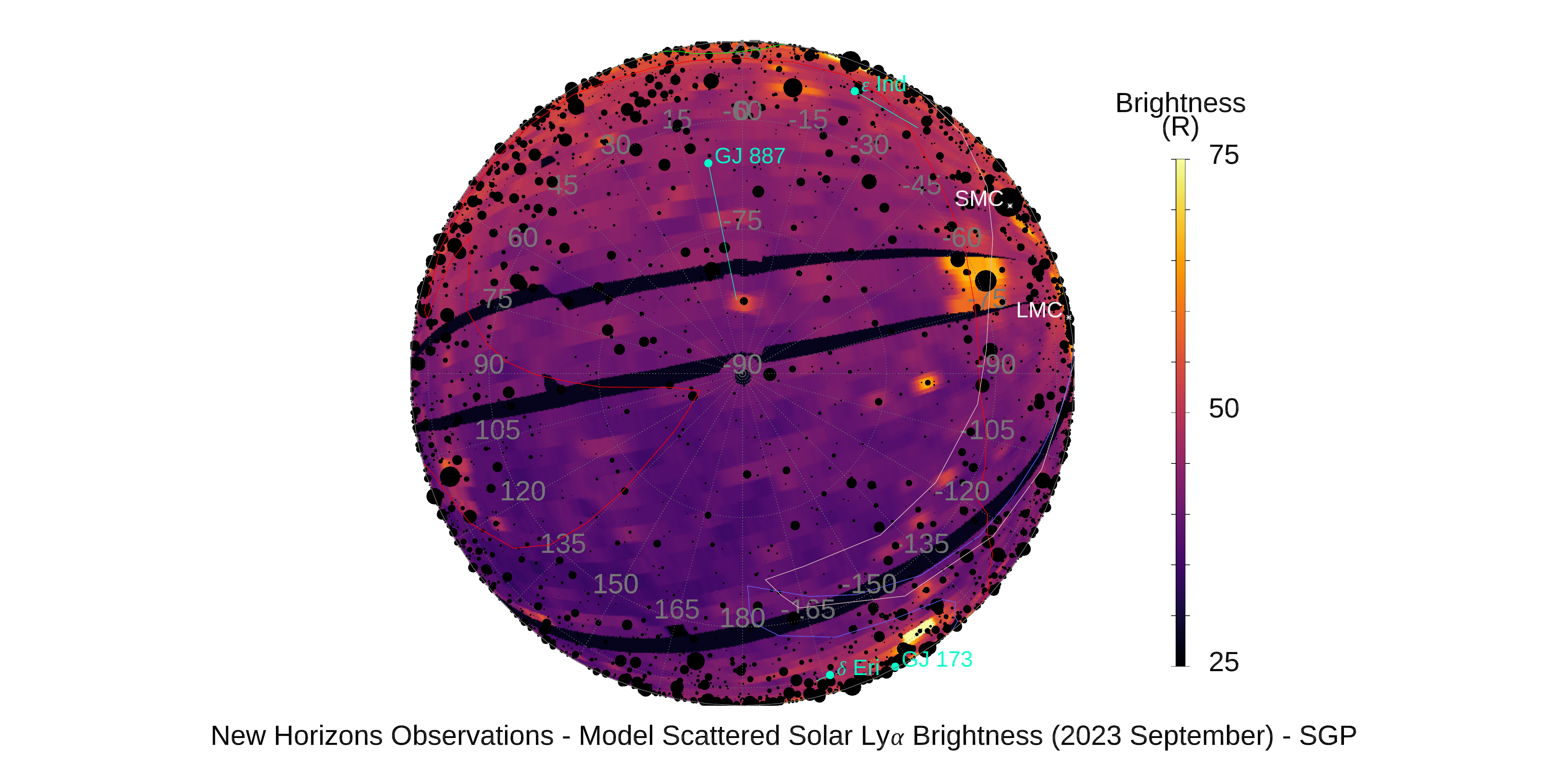}
\caption{Residual Ly$\alpha$ maps in Galactic coordinates, showing the polar regions (in satellite
projection) for the NGP (upper) and SGP (lower).
The typical diffuse brightness is $\sim5-10$~R larger in the north than in the south.
\label{fig:polarmaps}}
\end{figure}

Since the most useful data are at high Galactic latitudes, we also present residual Ly$\alpha$ maps for the 
regions of the north and south Galactic poles in Fig.~\ref{fig:polarmaps}.
Apart from some haloes surrounding a few bright stars, the polar regions are observed to be very uniform,
with only a small brightness gradient of $\sim 10$~R across $>60^\circ$ of latitude.
The north polar region is typically $\sim 10$~R brighter than the south polar region, i.e., $\sim 50$~R compared
to $\sim 40$~R.

We now consider what the sources of the diffuse Ly$\alpha$ are in the LISM, beginning with an overview
of how Ly$\alpha$ photons scatter in small and large LISM clouds.
The solar system is embedded in (or near) two small clouds (the LIC and the G clouds, Redfield \& Linsky 2008),
and one large cloud (the walls of the LHB).
Multiple scattering of Ly$\alpha$ photons inside a sufficently optically thick cloud that surrounds the solar
system would provide a simple way to make the diffuse emissions more isotropic, as observed in the NH maps.

\section{Resonance Line Radiative Transfer in the Local Interstellar Medium}

To understand where the Galactic Ly$\alpha$ emissions originate, it is important to understand how
Ly$\alpha$ photons scatter in the Local InterStellar Medium (LISM).
The sources of Ly$\alpha$ from recombination of \HI\ atoms following photoionization are only sufficient
to produce a few Rayleighs of brightness, if the Ly$\alpha$ photons were optically thin.
However, the brightness level inside a cloud (or as for the LHB, an empty shell) of \HI\ atoms can be
amplified by resonant scattering, if the cloud or shell is  sufficiently optically thick.
In this section we examine the scattering parameters of small clouds, such as the LIC or G clouds, and larger
structures, such as the walls of the LHB.
 
For a spherical cloud, with a Ly$\alpha$ optical depth of $\tau_{0}$ at line center, we can relate the number
of photon scatterings ($N_s$) and mean-free-path $\lambda_{\rm mfp}$ for line scattering opacity to the escape
of Ly$\alpha$ into the optically-thin damping wings.
The Doppler profile is characterized by a thermal velocity parameter
\begin{equation}
b = (2kT / m_{H})^{1/2} = (10.76~{\rm km~s}^{-1})~T_{7000}^{1/2}
\end{equation}
at $T = (7000~{\rm K})T_{7000}$ and the Doppler width of Ly$\alpha$ in frequency is
\begin{equation}
\Delta \nu_D = b \nu_0 / c = b / \lambda_0 \approx (8.85 \times 10^{10}~{\rm Hz}) T_{7000}^{1/2}.
\end{equation}
The intrinsic line width (Lorentzian width) is given by
\begin{equation}
\Delta \nu_L = A_{21} / {2\pi} \approx 9.97 \times 10^7~{\rm Hz},
\end{equation}
for the radiative decay rate of $A_{21} = 6.265 \times 10^8$~s$^{-1}$.
From these frequency widths the Voigt damping parameter (cf., Dijkstra 2014) is given by
\begin{equation}
a \equiv {{\Delta \nu_L} / {2 \Delta \nu_D}} = (5.63 \times 10^{-4}) T_{7000}^{-1/2}.
\end{equation}
We now switch to using the dimensionless frequency $x = (\nu - \nu_0) / \Delta \nu_D$ away from line center at
frequency $\nu_0$ and wavelength $\lambda = 121.567$~nm.
The line profile (Voigt function) can be approximated by
\begin{equation}
\phi (x) = \exp(-x^2)
\end{equation}
in the Doppler core and
\begin{equation}
\phi (x) \approx a / \sqrt{\pi} x^2
\end{equation}
in the damping wings. 
The dividing point between these regimes occurs at a dimensionless frequency $x_{crit} \approx 3.2$
for $T=7000$~K and $a / \sqrt{\pi} = 3.18 \times 10^{-4}$.
Diffusive scattering of Ly$\alpha$ emission lines in extremely optically thick media occurs at
large values of the product
\begin{equation}
a \tau_0 > 10^3.
\end{equation}
In this process, photons scatter repeatedly within the Doppler core of the line, and undergo a
random walk in frequency space into the wings.
The mean free path of a wing photon at frequency $x$ is
\begin{equation}
\lambda_{\rm mfp} \sim 1 / {\phi (x)}
\end{equation}
in units of line center optical depth $\tau_0$.
Photons in the line wing scatter $N_s \sim x^2$ times, usually returning to the core.
After many scatterings, the photons gradually shift out of the Doppler core into the optically thin
damping wings.
At that point, the photon mean free path becomes comparable to the cloud size, allowing photons to escape.

As the Ly$\alpha$ photon scatters repeatedly within the Doppler core, standard random-walk
arguments show that the photon moves to a dimensionless frequency in the damping wings,
$x_s \approx N_s^{1/2}$, in steps of the mean-free-path, $\lambda_{\rm mfp}$.
At this point, a photon will have diffused a ``distance'' $N_s^{1/2} / {\phi (x)}$
away from the cloud center (measured in line-center optical depths).
When we set this distance to the cloud line-center optical depth,
\begin{equation}
N_s^{1/2} / {\phi (x)} = \tau_0, 
\end{equation}
and use the approximation
\begin{equation}
\phi (x) \approx a / \sqrt{\pi} x^2
\end{equation}
appropriate for the damping wings, we find the scaling relation
\begin{equation}
x_s \approx (a \tau_0 / \sqrt{\pi})^{1/3}.
\end{equation}
center of any extremely opaque object, with two emission peaks in the line wings on either side of line-center
at frequencies 
\begin{equation}
x_p \approx \pm k (a \tau_0 / \sqrt{\pi})^{1/3},
\end{equation}
in dimensionless $\Delta \nu_D$ units, where $k \approx 1.1$ for a slab and $k \approx 0.92$ for a sphere.

Applying these expressions first to the small local clouds within 10~pc, with typical temperatures
$T \approx (7000~{\rm K}) T_{7000}$ and neutral H column densities $N_{\rm HI} = (10^{18}~{\rm cm}^{-2}) N_{18}$
(Redfield \& Linsky 2008, Linsky \& Redfield 2021), we find line-center optical depths
\begin{equation}
\tau_0 = \left({{\pi e^2} \over {m_e c}}\right) {{N_{\rm HI} f \lambda} \over {\sqrt{\pi} b}} \approx (70,440) N_{18} T_{7000}^{-1/2},
\end{equation}
where $b$ is as given above, the oscillator strength $f = 0.4164$, and $(\pi e^2 / m_e c) = 0.02654$~cm$^2$ s$^{-1}$.
The critical parameter 
\begin{equation}
a \tau_0 \approx 40 \ll 10^3
\end{equation}
for these small clouds, indicating that scattering into the damping wings is only partially complete.
The characteristic dimensionless scattering frequency would be
\begin{equation}
x_s \approx 2.8 N_{18}^{1/3} T_{7000}^{-1/6}.
\end{equation}
Thus, it seems that the small local clouds are not optically thick enough to isotropize any Ly$\alpha$ photons
incident on or created within them.

Now we consider the higher-density walls of the LHB.
If we parameterize the shell with \HI\ column density $N_{\rm HI} = (10^{20}~{\rm cm}^{-2}) N_{20}$, the same
formalism for Ly$\alpha$ scattering gives line-center optical depths
\begin{equation}
\tau_0 = (7.0 \times 10^6) N_{20} T_{7000}^{-1/2}. 
\end{equation}
The critical parameter for the LHB walls is 
\begin{equation}
a \tau_0 \approx 3940 > 10^3
\end{equation}
so that scattering into the damping wings is complete, and the characteristic dimensionless scattering frequency is
\begin{equation}
x_s \approx 13 N_{20}^{1/3} T_{7000}^{-1/6}.
\end{equation}
The neutral H at the LHB walls could be cooler than the assumed 7000~K, but the
$T^{-1/6}$ dependence of $x_s$ is weak.
The Ly$\alpha$ in the LHB walls is expected to be extremely optically
thick, approaching $10^7$ line scatterings out to dimensionless frequency $x_s \sim 10-20$ in the
line wings.
These lines will effectively be in the far-UV continuum near $\lambda \sim 121.6$~nm
and subject to dust absorption over length scales of a few hundred pc.
In the Zucker et~al. (2022) model, the estimated shell of displaced gas has a mass
$M_{sh} \approx 1.4 \times 10^6 M_\odot$ and average radius $R_{sh} \approx 165$~pc.
The mean (total hydrogen) column density would be
\begin{equation}
N_{\rm H} \approx {{M_{sh}} \over {4 \pi R_{sh}^2 (1.4 m_{H})}} \approx 3.7 \times 10^{20}~{\rm cm}^{-2},
\end{equation}
distributed over phases of \HI, H~II, and H$_2$.
They model the mean hydrogen density of the ambient (displaced) gas to be
$n_{\rm H} \approx 2.71_{-1.02}^{+1.57}$~cm$^{-3}$.
The large shell of H at the walls of the LHB does seem to be optically thick enough to isotropize
Ly$\alpha$ photons created within it.
 
Besides a large column of neutral H to scatter Ly$\alpha$ photons, an initial source of photons
is required.
As with early studies (e.g., M{\"u}nch~1962) we expect that most inital Ly$\alpha$ photons result
from recombination of regions photoionized by O and B stars.
Many such stars are found in the LHB (cf., Table~\ref{tab:lhbstars}) and there are several nearby clusters of
recently formed stars in and near the walls of the LHB.
Many of these star clusters appear toward one general direction, at Galactic longitudes of
$l \approx 353^\circ$ (Ophiuchus Cloud at $d \approx 140$~pc), $l \approx 351^\circ$
(Sco-Cen Association at $d \approx 130$~pc), and $l \approx 359^\circ$ (Corona Australis
Association at $d \approx 125$~pc).
Other potentially bright regions include the Taurus Association
($l \approx 174^\circ$, $d \approx 140$~pc), Pleiades Cluster ($l \approx 167^\circ$, $d \approx 135$~pc),
and Hyades Cluster ($l \approx 180^\circ$, $d \approx 46$~pc).
Much of the photoionization from these hot star clusters in the LHB wall is expected to result in
Ly$\alpha$ emission through recombination.
Based on the scattering arguments above, any Ly$\alpha$ that gets through the LHB wall and is detected
by the Alice spectrograph should be far enough into the line wings that no further resonant scattering within
the smaller nearby clouds would be expected.

\section{Discussion}

The residual Ly$\alpha$ maps presented earlier (Fig.~\ref{fig:resmaps} and Fig.~\ref{fig:polarmaps})
show mostly smooth variations at higher Galactic latitudes, with no apparent correlations with any
LISM structures that we had expected to see.
A summary of these absent structures is provided in Table~\ref{tab:structures}.
We believe that the lack of correlations and the general smoothness of the residual Ly$\alpha$
brightness map is best explained by resonant scattering of Ly$\alpha$ photons in the walls of the LHB. 
The Ly$\alpha$ brightness seen from the location of the solar system near the center of the
LHB are those photons which have escaped from some location on the interior wall of the LHB and were
passing through the bubble toward another location on the interior LHB wall. 
The initial source of the Ly$\alpha$ photons in the LHB walls is recombination of protons and electrons
following photoionization of neutral H atoms in the LHB walls by EUV radiation produced by hot stars
in the LHB and within its walls.

\begin{deluxetable*}{p{0.15\linewidth} p{0.3\linewidth} p{0.4\linewidth}}
\tablecaption{Absent Structures in Residual Ly$\alpha$ Map\label{tab:structures}}
\tablewidth{700pt}
\tabletypesize{\scriptsize}
\tablehead{\colhead{Feature} & \colhead{Lack of Expected Structure} & \colhead{Possible Explanation}}
\startdata
Hydrogen Wall &
No significant brightening in the direction of the heliopause, suggesting no detectable
hydrogen wall enhancement. &
A hydrogen wall could exist but be too faint and/or diffuse to be detected with the current datatset. \\
LISM Clouds &
No clear correlation between Ly$\alpha$ emissions and the expected density gradients of the LIC and G clouds. &
Ly$\alpha$ emissions from the wall of the LHB may be shifted by resonance scattering far enough from
line center that they aren't scattered by the LISM clouds. However, unresolved small-scale variations
in local cloud structure could still affect the Ly$\alpha$ brightness. Also, cloud boundary uncertainties
could affect the correlation analysis (cf., Gry \& Jenkins, 2014). \\
Fast-Moving Stars &
Absence of localized Ly$\alpha$ enhancements around high-velocity stars suggests minimal bowshock excitation. &
Bowshock Ly$\alpha$ emissions could be below the detection threshold or be present but unresolved
at the coarse spatial resolution of the Alice map. \\
\enddata
\end{deluxetable*}

Based on the above scattering arguments for distant Ly$\alpha$ emissions, the relatively isotropic 
nature of the residual Ly$\alpha$ map emissions indicates that they are likely produced in the nearby
LISM, specifically, in the walls of the LHB.
If the final emissions after multiple scattering are isotropic and homogeneous, the
radiative transfer equation becomes simpler.
Since the radiation doesn't change along any given line of sight and because scattering losses along that direction
are balanced by scattering gains from other directions, we can write (cf., Thomas \& Blamont 1976)
\begin{equation}
{J} \approx  k_{\rm abs} {4\pi I}
\end{equation}
where $J$ is the initial volume emission rate of Ly$\alpha$ photons in the nearby LHB walls, $4\pi I$ is the
Galactic background of $50$ R (i.e., $5.0 \times 10^7$~photons~cm$^{-2}$~s$^{-1}$), and $k_{\rm abs}$ is
the average coefficient for pure absorption by dust at $\lambda = 121.6$~nm, which is approximately
$6.9$~kpc$^{-1} = 2.24 \times 10^{-21}$~cm$^{-1}$ (from Draine~2011, and assuming a dust Ly$\alpha$
albedo of 0.5) for the LHB walls, using the parameters given above from Zucker et~al. (2022) and updated
in O'Neill et~al. (2024).
Thus, in order to produce the observed brightness of nearly isotropic Ly$\alpha$
emissions we need an initial Ly$\alpha$ volume emission rate of
$J \sim 5 \times 10^7 \times 2.24 \times 10^{-21} \sim 1.1 \times 10^{-13}$~photons~cm$^{-3}$~s$^{-1}$ or more.
While the ionization rate at the LHB walls will vary with location, a typical value might be what is measured
in the solar system.
Using the Velez et~al. (2024) stellar catalog and the O'Neill et~al. (2024) LHB wall distances, we find 54
stars that contribute most to the local ionization rate of $\Gamma_{\rm H I} \sim 6 \times 10^{-13}$~s$^{-1}$
(see Table~\ref{tab:lhbstars} of Appendix~B for details).
For an LHB wall neutral H density of $n_{\rm H I} \sim 3$~cm$^{-3}$, a typical photoionization rate
at the inside wall of the LHB would be $\sim 1.8 \times 10^{-12}$~cm$^{-3}$~s$^{-1}$.
In photochemical equilibrium, the initial Ly$\alpha$ volume emission rate would be $\sim 0.68$ of the photoionization 
rate or $J \sim 1.2 \times 10^{-12}$~photons~cm$^{-3}$~s$^{-1}$, which should be more than enough to produce the 
observed residual Ly$\alpha$ brightness.
Photoionization at the LHB walls from the hot gas within the bubble also contributes (cf., Fig.~2 of
Frisch et~al.~2011) but is neglected here since stellar photoionization dominates.
As noted in the analysis of the EUV spectrum of $\epsilon$~CMa by Shull et~al. (2025), for most early-type
stars there is a decrement of EUV flux near the Lyman limit at 91.1~nm, which may be inaccurately simulated
in the Kurucz model (LTE) fits to the observed FUV flux used in the Velez et~al. (2024) catalog.
This would mean that our estimated photoionization rate could be substantially underestimated.

It is important to note that Ly$\alpha$ photons undergoing resonance line radiative transfer in the walls
of the LHB are able to move freely back and forth across the cavity inside the walls without being absorbed
or scattered by the hot ionized gas in the LHB (although each crossing will take several hundred years). 
Thus, inside the LHB at the location of the solar system, the local Ly$\alpha$ is amplified by multiple
scattering so that what would provide an optically thin brightness of $\sim1$~R is increased to an optically
thick value of $\sim50$~R within the LHB.
We have run simple `slab' plane-parallel models of the LHB walls, using the angle-averaged partial frequency redistribution
resonance scattering code of Gladstone (1982) with a midplane initial volume emission rate and LHB wall model consistent
with the above estimates, and confirm that an amplification factor of $\sim 50$ is relatively easily to achieve
(while similar models for the LIC and G clouds are unable to produce an amplification factor of more than a few).
These 1-D slab models are not realistic enough to confirm our suggested source for the observed diffuse
Ly$\alpha$ as being due to resonance scattering in the LHB walls of Ly$\alpha$ photons produced by recombination following
photoionization by O and B stars within the LHB, but they strongly point to it as a likely explanation.

These models indicate that the expected line profile of the diffuse Ly$\alpha$ galactic emissions observed from
the solar system would be flat-topped (cf., internal line profiles, as in Meier \& Lee 1981) with a full width of 
$\pm 0.016$~nm, about twice as broad as the backscattered solar line width.
Any instruments capable of resolving the Ly$\alpha$ line profile (e.g., HST/STIS, MAVEN/IUVS) should expect to see a
Galactic background component of roughly $50\pm20 / 0.016 \sim 3125\pm1250$~R~nm$^{-1}$ over this wavelength interval.
More sophisticated 3-D radiative transfer calculations are required to make further progress, but
it seems plausible that the residual Ly$\alpha$ maps produced here (Fig.~\ref{fig:resmaps} and Fig.~\ref{fig:polarmaps})
may provide a new way to investigate the interior walls of the LHB.
As a start in this direction, in Appendix~B we show how the stellar photoionization rate would vary across the LHB wall,
and we provide a map of the expected H$\alpha$ brightness due to recombination of neutral H in the LHB walls.
This map should be similar to what the Ly$\alpha$ brightness would look like if the emissions were optically
thin (instead of the expected value of $\sim1.3 \times 10^7$ optical depths at line center).

Although our explanation for the observed properties of the residual Ly$\alpha$ map seems plausible,
there are several key assumptions that are critical for it to work, some of which are presented
in Table~\ref{tab:assumptions}.
We expect that the simplest way to make further progress is to develop a 3-D resonance line radiative
transfer code which incorporates the known structure of the LHB, its walls, the local clouds within it,
and the hot stars within it.

\begin{deluxetable*}{p{0.15\linewidth} p{0.3\linewidth} p{0.4\linewidth}}
\tablecaption{Important Assumptions\label{tab:assumptions}}
\tablewidth{700pt}
\tabletypesize{\scriptsize}
\tablehead{ \colhead{Key Assumption} & \colhead{Description} & \colhead{Potential Limitations}}
\startdata
Calibration and Sensitivity of Alice Spectrograph &
Alice's sensitivity calibration is assumed to be accurate for Ly$\alpha$ brightness conversion. &
Background subtraction uncertainties and potential instrumental degradation over time may affect
absolute brightness estimates. \\
Hall Model &
This model accounts for backscattered solar Ly$\alpha$ which is subtracted from the observed
Ly$\alpha$ brightness to produce the residual Ly$\alpha$ map. &
The Hall model neglects charge exchange reactions, and so does not include a hydrogen wall.
It also includes the simplifying but inaccurate assumptions of complete frequency redistribution
in its resonant scattering calculations, and cylindrical symmetry about the upstream/downstream axis. \\
Uniformity of Residual Ly$\alpha$ Emission &
Residual Ly$\alpha$ brightness ($50\pm20$~R) is interpreted as a uniform background from recombination
in LHB walls, followed by resonant scattering. &
Localized variations in ionizing sources or extragalactic contributions could introduce small-scale structure.
A 3-D radiative transfer model is needed. \\
Resonant Scattering and Optical Depth &
The LHB walls are assumed to be highly optically thick at line center ($\tau \sim 10^7$), ensuring
multiple scattering that can isotropize Ly$\alpha$. &
Small-scale density variations or turbulence in the LHB walls could introduce anisotropies in
Ly$\alpha$ emission. A 3-D radiative transfer model is needed. \\
\enddata
\end{deluxetable*}

\section{Conclusions}

The New Horizons Alice instrument has been used to obtain the first detailed all-sky map of Ly$\alpha$ emission 
observed from the outer solar system, where the Galactic and solar contributions to the observed brightness are comparable, 
and the solar contribution can be reasonably removed.
The residual Ly$\alpha$ map can then be searched for structures (i.e., sources and sinks) of emissions in the 
heliosphere and LISM.
Near the heliosphere, a brightness enhancement due to a hydrogen wall at the heliopause 
(e.g., Qu{\'e}merais et~al.~2010; Izmodenov et~al.~2013) is not seen, but it is expected to be very large scale and 
could be present at a $\sim10$~Rayleigh level.
No apparent correlations are seen with the boundaries of the four main clouds of the LISM (e.g., Redfield \& Linsky~2008);
as with the hydrogen wall, these could exist at a $\sim10$~Rayleigh level.
Apart from its structure on the $10-20$~R level, the overall average brightness level of the \Lya\ emission of $\sim50$~R
can be explained as largely due to early-type stars (and hot gas) in the Local Hot Bubble shining on the interior walls
of the LHB and photoionizing the \HI\ there.
The recombining protons produce Ly$\alpha$ (and concomitant H$\alpha$ and two-photon continuum) emissions, which are resonantly
scattered in the walls of the LHB --- and across the LHB cavity --- until they are approximately isotropic.
Since structures like the LHB are fairly common features of galaxies, it may be that the
interiors of similar bubbles elsewhere could also be regions of enhanced diffuse Ly$\alpha$ emissions.

A follow-up NH Alice all-sky Ly$\alpha$ map may be made in the future, if possible, and combining that
map with this map could result in a considerable improvement in angular resolution.
Finally, the maps presented here were obtained using the Alice spectrograph as a photometer,
since its spectral resolution is too coarse to resolve the details of the Ly$\alpha$ line structure.
However, there are instruments capable of resolving the Ly$\alpha$ line profile (e.g., Clarke et~al.~1998; Hosseini \& Harris~2020)
which could possibly study this emission in more detail, and thus (even from Earth orbit) provide a new window on the LISM and
H populations in the heliosphere.

\section{Appendix~A: Sensitivity of Alice to Diffuse Ly$\alpha$}

The all-sky Ly$\alpha$ maps presented here depend on an accurate conversion of the observed Alice total analog count rate 
into Ly$\alpha$ brightness.
In a previous paper (Gladstone et~al.~2021) we found a conversion factor, or sensitivity, of $4.92\pm0.09$~counts s$^{-1}$ R$^{-1}$.
However, we now believe that value is too large, because we mistakenly used the {\it digital} background count rate instead
of the {\it analog} background count rate in deriving the sensitivity.
Since the Alice analog count rate is substantially larger than its digital count rate (typically $\sim170$~counts s$^{-1}$ compared
to $\sim120$~counts s$^{-1}$), the subtracted background was underestimated in (Gladstone et~al.~2021), and the derived sensitivity
of $4.92\pm0.09$~counts s$^{-1}$ R$^{-1}$ is too large.

This error is corrected here, with the help of several Alice observations made in support of a study of the Cosmic Optical
Background (COB), at about the same time period as the all-sky Ly$\alpha$ map, using the LORRI instrument on New Horizons.
As reported by Postman et~al. (2024), fifteen observations of the COB were made in a variety of directions (chosen to have
as little stellar flux as possible), at high Galactic latitudes.
The directions and epochs of the supporting Alice observations are listed in Table~\ref{tab:ncob}, along with the measured Ly$\alpha$ brightness
in the narrow slit and the total analog Alice count rate.
Each Alice observation consisted of eight 1-hour histogram-mode exposures, interleaved with eight 1-hour dark exposures, providing
a very deep FUV spectral image in each of the fifteen directions.  

\begin{deluxetable*}{lccccccccccc}
\tablecaption{New Horizons NCOB Observations of Specific Sky Directions\label{tab:ncob}}
\tablewidth{700pt}
\tabletypesize{\scriptsize}
\tablehead{
\colhead{Observation$^a$} & \colhead{R.A.} & \colhead{Dec.} & \colhead{Roll$^b$} &
 \colhead{$l$} & \colhead{$b$} & \colhead{B(Ly$\alpha$)} &
 \colhead{Analog Rate} & \colhead{$d_{\rm SUN}$} & \colhead{Start Date} & \colhead{Start MET} & \colhead{End MET} \\
 & ($^\circ$) & ($^\circ$) & ($^\circ$) & ($^\circ$) & ($^\circ$) & (R) & (counts s$^{-1}$) & (AU) & (UT) & (s) & (s)
 } 
\startdata
NCOB01 & 358.4334 & -54.9137 & 90 & 319.728 & -60.293 & $136.7\pm4.7$ & $485.7\pm0.9$ & 56.97 & 2023-09-13 13:56:00 & 556919276 & 556981541 \\
NCOB02 & 5.3540 & -55.6590 & 180 & 311.662 & -60.958 & $131.5\pm4.7$ & $484.3\pm0.9$ & 56.86 & 2023-08-30 18:24:51 & 555725807 & 555788037 \\
NCOB03 & 353.7867 & -49.1893 & 90 & 331.516 & -63.490 & $133.4\pm4.7$ & $490.5\pm0.9$ & 56.79 & 2023-08-21 15:05:51 & 554936267 & 554994836 \\
NCOB04 & 8.0987 & -44.4906 & 135 & 314.112 & -72.222 & $139.0\pm4.7$ & $499.5\pm0.9$ & 56.78 & 2023-08-20 21:46:31 & 554873907 & 554932442 \\
NCOB05 & 10.7611 & -27.3461 & 180 & 25.794 & -88.122 & $131.9\pm4.7$ & $491.7\pm0.9$ & 56.84 & 2023-08-28 14:49:31 & 555540087 & 555598622 \\
NCOB06 & 9.4350 & -34.7328 & 180 & 323.189 & -81.850 & $136.5\pm4.7$ & $489.2\pm0.9$ & 56.96 & 2023-09-12 11:41:40 & 556824816 & 556883342 \\
NCOB07 & 19.0398 & -26.6161 & 180 & 209.027 & -84.464 & $139.0\pm4.8$ & $507.5\pm1.0$ & 56.84 & 2023-08-27 21:42:31 & 555478467 & 555537002 \\
NCOB08 & 336.2651 & -30.0473 & 180 & 18.925 & -57.870 & $125.2\pm4.6$ & $456.8\pm0.9$ & 56.86 & 2023-08-30 01:09:31 & 555663687 & 555722222 \\
NCOB09 & 6.7398 & -22.1689 & 180 & 73.341 & -82.551 & $133.5\pm4.7$ & $489.5\pm0.9$ & 56.85 & 2023-08-29 07:56:31 & 555601707 & 555660242 \\
NCOB10 & 15.7115 & -18.8994 & 180 & 141.203 & -81.364 & $137.9\pm4.8$ & $502.9\pm1.0$ & 56.83 & 2023-08-27 04:36:31 & 555416907 & 555475442 \\
NCOB11 & 10.6266 & -15.2837 & 180 & 112.541 & -77.975 & $133.9\pm4.7$ & $496.5\pm1.0$ & 56.83 & 2023-08-26 06:16:01 & 555336477 & 555413182 \\
NCOB12 & 207.4692 & 3.9649 & 0 & 336.539 & 62.959 & $150.6\pm4.9$ & $556.5\pm1.0$ & 56.75 & 2023-08-17 04:02:00 & 554550836 & 554613057 \\
NCOB13 & 211.9528 & 4.6995 & 0 & 345.372 & 61.111 & $148.2\pm4.8$ & $548.9\pm1.0$ & 56.75 & 2023-08-16 10:51:30 & 554489006 & 554547532 \\
NCOB14 & 356.2651 & 15.5111 & 135 & 100.270 & -44.420 & $134.2\pm4.7$ & $501.7\pm1.0$ & 56.78 & 2023-08-19 19:27:59 & 554779195 & 554837722 \\
NCOB15 & 247.9273 & 55.2059 & 0 & 84.133 & 41.702 & $140.2\pm4.8$ & $530.0\pm1.0$ & 56.72 & 2023-08-12 12:56:40 & 554150916 & 554209442 \\
\enddata
\noindent $^a$ As reported in Postman et~al. (2024). \\
\noindent $^b$ Alice slit position angle, measured counterclockwise from the north celestial pole. \\
\end{deluxetable*}

\begin{figure}[ht]
\centering
\includegraphics[width=6.5in]{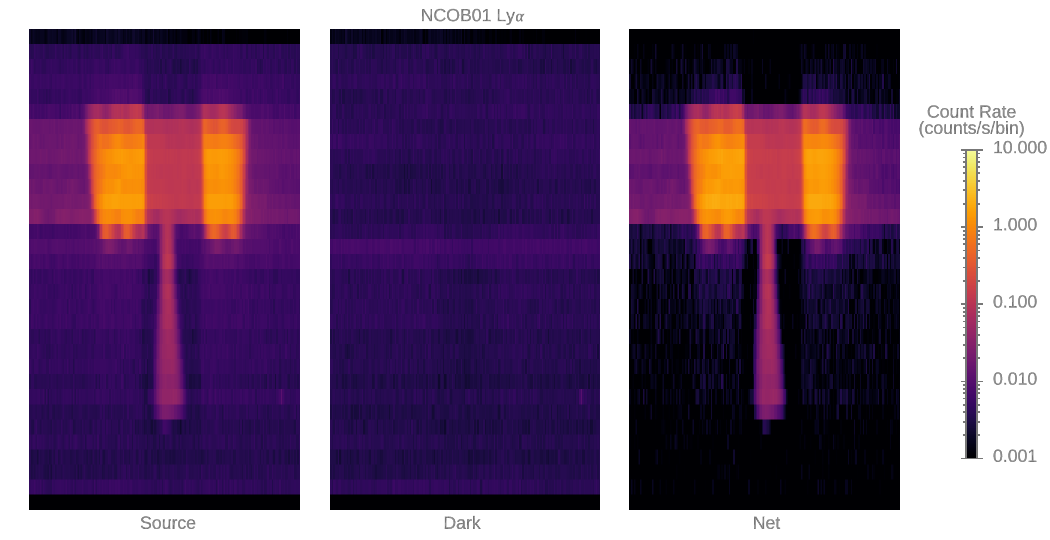}
\caption{Spectral images for the Alice observation of a dark region on the sky, referred to 
as NCOB01 (see Postman et~al. (2024) and Table~\ref{tab:ncob} for more information).
{\it Left:} The count rate spectral image of the sum of eight 1-hour histogram exposures of the
sky, using all 32 rows and columns 535 through 555 of the Alice detector in order to display the 
spectral image in the Ly$\alpha$ vicinity.   Approximate wavelengths are 120.3-123.9~nm; 
 most of the signal is from diffuse Ly$\alpha$ filling the Alice slit.
{\it Center:} Same as Left, but for the sum of the eight 1-hour interleaved dark exposures for 
observations NCOB01.
{\it Right:} The net count rate of the sky (Left) after subtracting the dark (Center) spectral images, 
showing that the dark sky is completely dominated by diffuse Ly$\alpha$ at FUV wavelengths.
The count rate is shown on a log scale in the color bar on the right.
\label{fig:specimage}}
\end{figure}

An example spectral image (with accompanying dark) is shown in Fig.~\ref{fig:specimage}.
The spectral images are truncated in the spectral direction (i.e., the x-direction) to show only the region around the
Ly$\alpha$ line, but include all 32~rows of the Alice detector.
Each detector row subtends $0.27^\circ$ on the sky, and the relationship between off-axis angle in the spectral direction and
wavelength is $10$~nm degree$^{-1}$, so that the spectral images display the Alice $2^\circ\times2^\circ+0.1^\circ\times4^\circ$ slit
in the approximately correct geometric shape.

\begin{figure}[ht]
\centering
\includegraphics[width=6.5in]{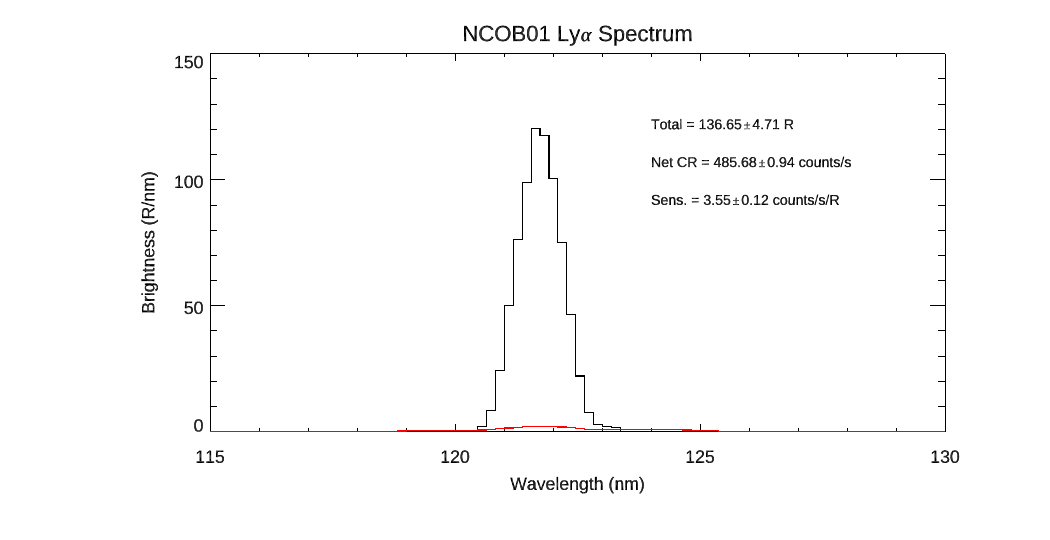}
\caption{An example spectrum of wavelengths including the diffuse Ly$\alpha$ from the NCOB01 deep exposure, 
comprised of eight 1-hour exposures interleaved with eight 1-hour darks with the Alice instrument pointed at
the direction $(l,b) = (319.728^\circ,-60.293^\circ)$ (see Table~\ref{tab:ncob} for other NCOB information).
The observed net brightness spectrum is given by the black histogram, and the 1-$\sigma$ error of the net
brightness is shown by the red histogram.
As noted on the figure, for this observation the on-axis average diffuse Ly$\alpha$ brightness was
$136.65\pm4.71$~R, and the corresponding analog net count rate was $485.68\pm0.94$~counts s$^{-1}$.
These values are combined to give a Ly$\alpha$ sensitivity for the Alice spectrograph of 
$3.55\pm0.12$~counts s$^{-1}$ R$^{-1}$ for this observation. 
\label{fig:ncoblya}}
\end{figure}

\begin{figure}[ht]
\centering
\includegraphics[width=6.5in]{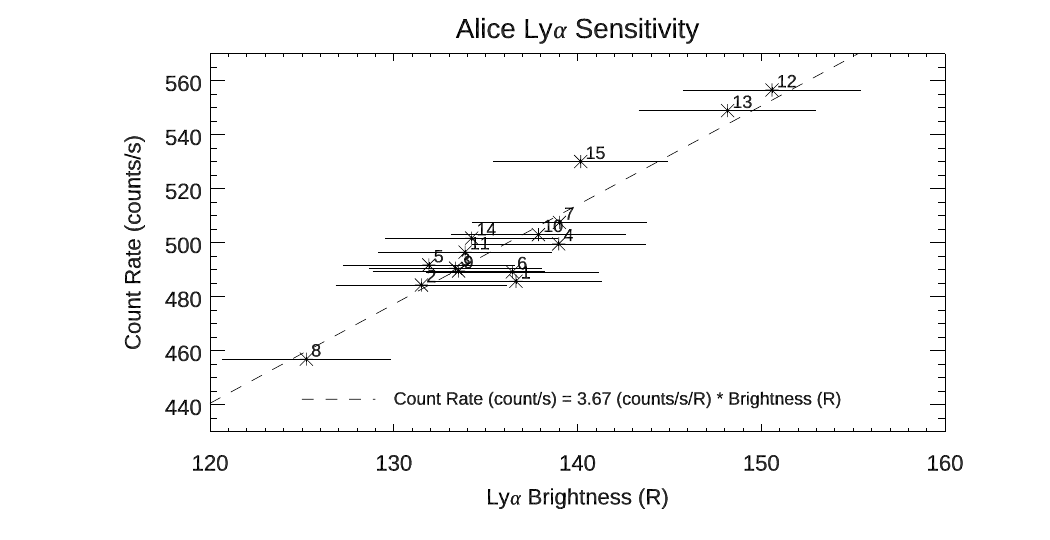}
\caption{The variation of the Alice spectrograph total analog count rate to observed diffuse Ly$\alpha$
brightness is plotted for the fifteen NCOB observations.
Using the 15~NCOB measurements we find an average sensitivity for the Alice spectrograph of $3.67\pm0.02$~counts s$^{-1}$ R$^{-1}$, 
which is overplotted as a dashed line.
This sensitivity is used to convert the analog count rates observed during the all-sky map into diffuse Ly$\alpha$
brightnesses in regions of the sky uncontaminated by UV-bright stars.
\label{fig:sensitivity}}
\end{figure}

Since the Alice instrument is calibrated with boresight (row 16) observations of standard stars, we use spectra
of that row to determine the integrated Ly$\alpha$ brightness for each of the NCOB observations, e.g.,
as shown in Fig.~\ref{fig:ncoblya} for NCOB1.
Finally, we compare all fifteen NCOB observations in Fig.~\ref{fig:sensitivity}, the average of which
yields a sensitivity of the Alice spectrograph of $3.67\pm0.02$~counts s$^{-1}$ R$^{-1}$.

\section{Appendix~B: Constraints on H$\alpha$ Produced in the Walls of the Local Bubble}

As argued above, the residual all-sky Ly$\alpha$ brightness of $50 \pm 20$~R is likely produced in
the walls of the Local Hot Bubble (LHB).
Sources of Ly$\alpha$ more distant than the outside wall of the LHB would be expected to be less isotropic, i.e.,
Ly$\alpha$ photons from more distant sources would be more likely to get through the LHB wall if they
had already scattered in wavelength into the line wings, and would subtend a smaller solid
angle than nearer sources.

Here we estimate the contribution to the observed H$\alpha$ from recombination of H atoms which
have been photoionized by EUV emission from stars within the outer wall of the LHB, in the same process
which leads to Ly$\alpha$ emission.
This predicted LHB wall H$\alpha$ emission can then be compared with maps of H$\alpha$ (e.g., Finkbeiner~2003) to
search for any correlation.
The dust absorption coefficient for H$\alpha$ photons is about one-sixth as large as for Ly$\alpha$ photons
(Draine~2011), so that most of the H$\alpha$ signal comes from sources far beyond the LHB wall, and the
contribution from the LHB wall should be no more than a fraction of the total observed signal.
We take the LHB to be hot and highly ionized (e.g., Snowden et~al.~2014; Slavin et~al.~2017), so that it is
optically thin to EUV photons.  
At $T=10^6$~K, photochemical equilibrium indicates the \HI\ density expected for a balance between
collisional ionization and recombination would be $n_{\rm HI} \approx 10^{-8}$~cm$^{-3}$ or less, since the 
recombination rate coefficient decreases rapidly at high temperatures.
Thus, the EUV photons produced by hot O and B stars in the LHB will first be absorbed when they reach the 
walls of the LHB.
The \HI\ column density in the LHB walls is large ($N_{\rm HI} > 10^{20}~{\rm cm}^{-2}$, with 
an estimated density $n_{\rm HI} \approx 3$~cm$^{-3}$, Zucker et~al.~2022).
So we would expect most the EUV flux from the hot LHB stars to photoionize neutral H in the LHB walls.
About half of recombinations (actually 0.452, for case-B recombination at $T = 10^4$~K as given by Draine~2011)
following these would lead to an approximately equivalent flux of H$\alpha$ photons.

\begin{figure}[ht]
\centering
\includegraphics[width=6.5in]{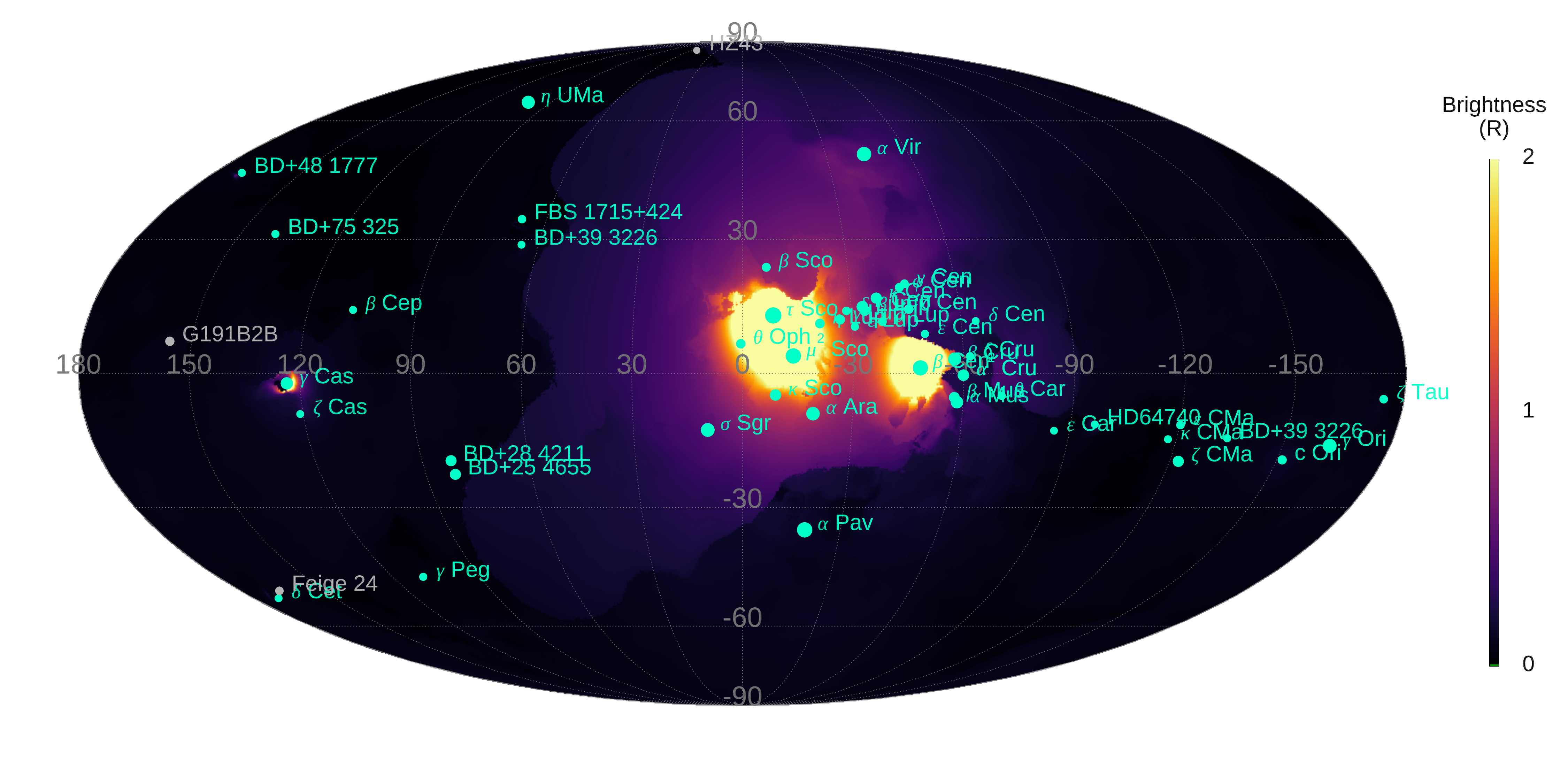}
\includegraphics[width=6.5in]{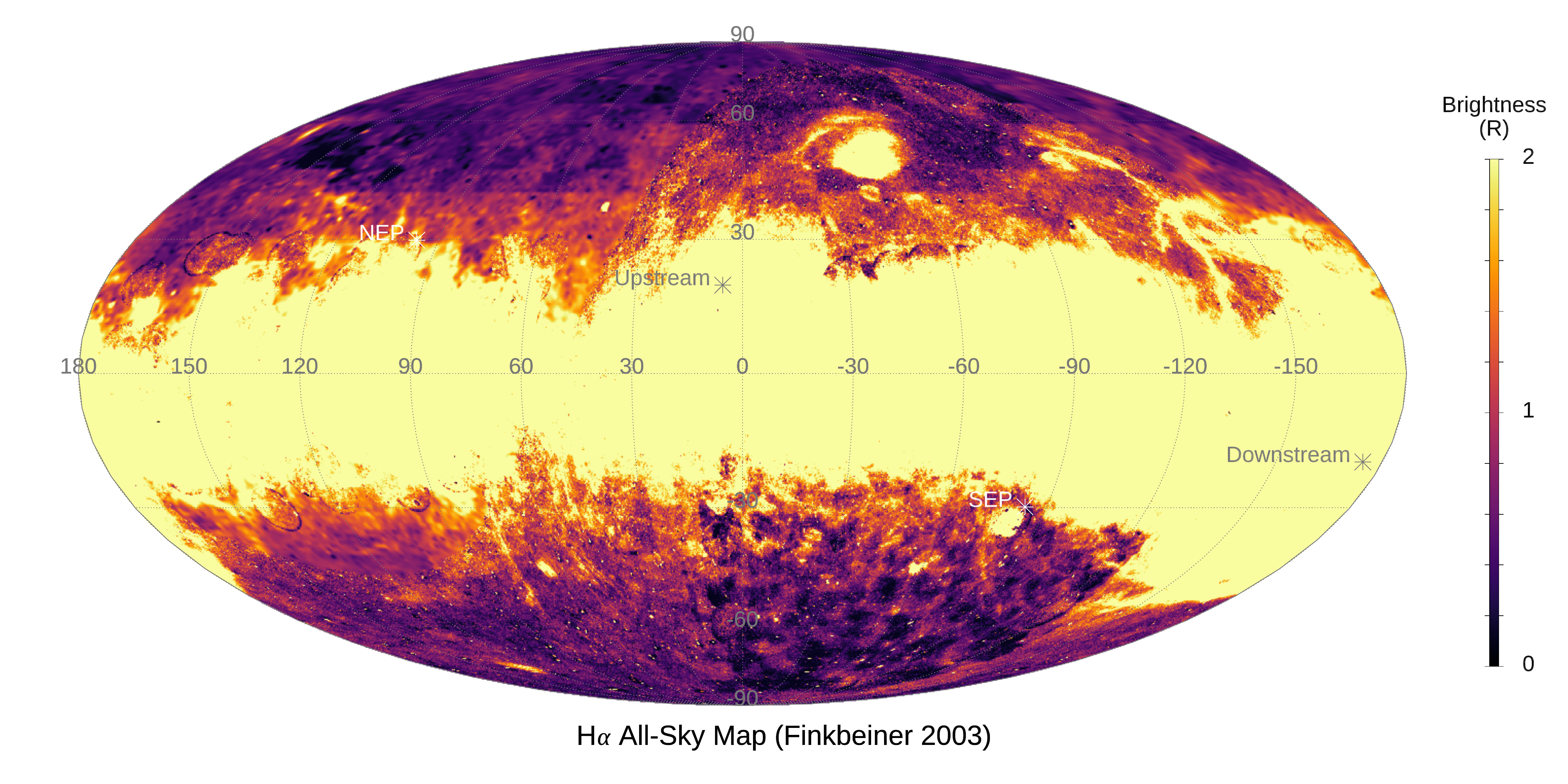}
\caption{The upper panel shows the estimated brightness of H$\alpha$ emissions from the wall of the LHB that would
result from EUV illumination by hot stars within the LHB, followed by local recombination and emission.
The location of the outer edge of the LHB, as provided in O'Neill et~al. (2024), was used in this calculation,
along with the radiation field estimated using the stars in Table~\ref{tab:lhbstars}, which are labelled with dots sized according
to the logarithm of their estimated \HI\ photoionizing flux at a distance of $1$~pc.
The brightest regions are near the primary O-B associations mentioned above at the end of section~4.
The lower panel shows a high-resolution map of diffuse H$\alpha$ emissions (Finkbeiner~2003); the lack of
correlation with the model map of expected H$\alpha$ from the interior wall of the LHB is likely due to 
the domination of contributions from beyond the outer wall of the LHB.
\label{fig:halpha}}
\end{figure}

Using the Velez et~al. (2024) catalog we find 51 O and B stars in the LHB or the LHB wall (i.e., their distances
were less than the distance to the outer LHB wall in the direction of the star, as estimated by O'Neill et~al. (2024).
To this list we added three white dwarfs (G191B2B, Feige 24, and HZ43) known to be important sources of EUV
radiation near the solar system (e.g., Dupuis et al.~1995; Welsh et~al.~2013).
These 54 stars are listed in Table~\ref{tab:lhbstars}, along with their salient properties.
We estimated the EUV flux from these stars at each location on the LHB wall, then estimated the H$\alpha$
brightness map that source would produce locally at the LHB wall as seen from Earth, which is shown in Fig.~\ref{fig:halpha}.
Since the recombination yields H$\alpha$ and Ly$\alpha$ are similar, this map effectively shows how the
residual all-sky Ly$\alpha$ map shown in Fig.~\ref{fig:resmaps} would appear if the Ly$\alpha$ photons produced
by recombination in the LHB walls were optically thin.
As seen in Fig.~\ref{fig:halpha}, the brightness expected from this source of diffuse H$\alpha$ from the LHB walls is
generally only a Rayleigh or less, and it doesn't contribute much to the total sky brightness of H$\alpha$ (e.g.,
Finkbeiner~2003), which is shown on the same brightnesss scale for comparison.
We note that two-photon continuum emissions are also a result of the recombination process (e.g., Kulkarni~2021);
these should be produced at roughly one-half of the production rate of Ly$\alpha$ emissions, and would also expected
to be at the $\sim$ 1~R level when integrated over wavelength.
Future 3-D studies of multiple resonant scattering of Ly$\alpha$ photons in the LHB walls may find this map useful
for comparing with their initial Ly$\alpha$ emission rate.

\begin{deluxetable*}{lcclcccccc}
\tablecaption{Brightest EUV Stars in the Local Hot Bubble\label{tab:lhbstars}}
\tablewidth{700pt}
\tabletypesize{\scriptsize}
\tablehead{
\colhead{Name$^a$} & \colhead{\HI\ Ionization} & \colhead{\HI\ Ionizing} & \colhead{Spectral} &
 \colhead{$l$} & \colhead{$b$} & \colhead{Stellar} & \colhead{Outer LHB} & $\log N_{\rm HI}$$^e$ & n$_{\rm HI}$$^e$ \\
  & Rate$^b$ & Flux$^c$ & Type & & & Distance & Wall Distance$^d$ & & \\
 & (s$^{-1}$) & (photons cm$^{-2}$ s$^{-1}$) & & ($^\circ$) & ($^\circ$) & (pc) & (pc) & ($N$ in cm$^{-2}$) & (cm$^{-3}$)
 } 
\startdata
$\tau$ Sco      & $1.43\times10^{-13}$ & $7.52\times10^8$ & B0.2V & 351.54 & 12.81 & 145.3 & 164.3 & 20.5 & 0.7 \\
$\mu^2$ Sco     & $7.60\times10^{-14}$ & $5.75\times10^8$ & B2IV & 346.20 & 3.86 & 176.6 & 205.0 & & \\
$\beta$ Cen	& $6.86\times10^{-14}$ & $2.27\times10^8$ & B1III & 311.77 & 1.25 & 120.2 & 141.4 & 19.63 & 0.12 \\
$\alpha$ Pav	& $7.91\times10^{-14}$ & $5.43\times10^7$ & B2IV & 340.90 & -35.19 & 54.8 & 291.6 & & \\
$\alpha$ Vir	& $4.09\times10^{-14}$ & $5.39\times10^7$ & B1V & 316.11 & 50.85 & 76.6 & 129.4 & $<19.0$ & $<0.04$ \\
$\gamma$ Cas	& $7.84\times10^{-15}$ & $4.92\times10^7$ & B0.5IVpe & 123.58 & -2.15 & 168.4 & 273.9 & & \\
$\gamma$ Ori	& $2.45\times10^{-14}$ & $3.56\times10^7$ & B2V & 196.93 & -15.95 & 77.4 & 170.0 & & \\
$\alpha$ Ara 	& $2.12\times10^{-14}$ & $3.18\times10^7$ & B2Vne & 340.76 & -8.83 & 82.0 & 195.6 & & \\
$\sigma$ Sgr	& $2.21\times10^{-14}$ & $2.68\times10^7$ & B2V & 9.56 & -12.44 & 69.8 & 166.3 & & \\
$\beta$ Cru	& $1.57\times10^{-14}$ & $2.60\times10^7$ & B1IV & 302.46 & 3.18 & 85.4 & 138.1 & & \\
$\kappa$ Sco    & $4.05\times10^{-15}$ & $1.97\times10^7$ & B1.5III & 351.04 & -4.72 & 148.1 & 157.6 & 20.2 & 0.35 \\
$\alpha$ Mus	& $9.30\times10^{-15}$ & $1.92\times10^7$ & B2IV & 301.66 & -6.30 & 96.7 & 155.1 & $<20.4$ & $<0.8$ \\
FBS 1715+424    & $3.13\times10^{-16}$ & $1.73\times10^7$ & sd0He0 & 67.50 & 34.75 & 422.2 & 574.0 & & \\
$\kappa$ Cen	& $2.86\times10^{-15}$ & $1.22\times10^7$ & B2IV & 326.87 & 14.75 & 117.5 & 191.6 & $<20.6$ & $<1.1$ \\
$\alpha^2$ Cru	& $4.43\times10^{-15}$ & $9.82\times10^6$ & B1V & 300.13 & -0.36 & 99.0 & 207.6 & & \\
$\zeta$ CMa	& $2.48\times10^{-15}$ & $9.15\times10^6$ & B2.5V & 237.52 & -19.43 & 111.1 & 284.9 & & \\
BD+25 4655 	& $2.29\times10^{-15}$ & $8.96\times10^6$ & sdO6 & 81.67 & -22.36 & 115.5 & 230.1 & & \\
$\alpha^1$ Cru	& $3.85\times10^{-15}$ & $8.62\times10^6$ & B0.5IV & 300.13 & -0.36 & 99.0 & 207.6 & & \\
BD+28 4211 	& $2.21\times10^{-15}$ & $8.62\times10^6$ & sdO2VIIIHe5 & 81.87 & -19.29 & 112.1 & 215.3 & & \\
$\beta$ Lup	& $2.61\times10^{-15}$ & $7.38\times10^6$ & B2III & 326.25 & 13.91 & 117.4 & 191.3 & 20.4 & 0.7 \\
$\eta$ Cen	& $3.46\times10^{-15}$ & $6.72\times10^6$ & B2Ve & 322.77 & 16.67 & 93.7 & 192.9 & & \\
$\beta$ Mus	& $2.70\times10^{-15}$ & $6.54\times10^6$ & B2V+B3V & 302.45 & -5.24 & 104.7 & 155.5 & & \\
$\gamma$ Lup	& $1.35\times10^{-15}$ & $4.60\times10^6$ & B2IV & 333.19 & 11.89 & 129.0 & 189.9 & $<20.3$ & $<0.5$ \\
$\eta$ Lup	& $1.18\times10^{-15}$ & $4.21\times10^6$ & B2.5IV & 338.77 & 11.01 & 131.7 & 185.1 & & \\
BD+48 1777 	& $3.43\times10^{-16}$ & $3.89\times10^6$ & sdO9: & 170.34 & 46.04 & 236.1 & 335.5 & & \\
c Ori	        & $7.22\times10^{-16}$ & $3.76\times10^6$ & B1V & 208.50 & -19.11 & 153.5 & 174.5 & & \\
HD64740	        & $2.94\times10^{-16}$ & $3.61\times10^6$ & B2V & 263.38 & -11.19 & 244.8 & 277.7 & & \\
$\theta$ Oph	& $9.84\times10^{-16}$ & $3.59\times10^6$ & B2IV & 0.46 & 6.55 & 133.7 & 166.8 & $<20.5$ & $<0.8$ \\
$\eta$ UMa	& $1.29\times10^{-14}$ & $3.38\times10^6$ & B3V & 100.70 & 65.32 & 31.9 & 361.6 & $<21.0$ & $<10$ \\
$\zeta$ Cen	& $9.85\times10^{-16}$ & $2.80\times10^6$ & B2.5IV & 314.07 & 14.19 & 117.7 & 189.9 & & \\
$\beta$ Cep	& $3.05\times10^{-16}$ & $2.76\times10^6$ & B0.5IIIs & 107.54 & 14.03 & 210.1 & 291.4 & 19.93 & 0.13 \\
$\kappa$ CMa	& $3.13\times10^{-16}$ & $2.76\times10^6$ & B1.5Ve & 242.36 & -14.49 & 207.0 & 270.7 & & \\
$\phi$ Cen	& $6.83\times10^{-16}$ & $2.76\times10^6$ & B2IV & 315.98 & 19.07 & 140.8 & 189.7 & & \\
$\alpha$ Lup	& $6.19\times10^{-16}$ & $2.51\times10^6$ & B1.5III & 321.61 & 11.44 & 142.5 & 209.6 & $<19.3$ & $<0.04$ \\
$\delta$ Cet	& $3.10\times10^{-16}$ & $2.42\times10^6$ & B2IV & 170.76 & -52.21 & 194.9 & 258.0 & & \\
$\delta$ Lup	& $5.14\times10^{-16}$ & $2.32\times10^6$ & B1.5IV & 331.32 & 13.82 & 148.6 & 189.0 & $<20.2$ & $<0.4$ \\
BD+39 3226 	& $2.92\times10^{-16}$ & $2.13\times10^6$ & sdOHe & 65.00 & 28.77 & 189.2 & 281.3 & & \\
$\beta$ Sco     & $6.35\times10^{-16}$ & $2.02\times10^6$ & B1V+B2V & 353.19 & 23.60 & 120.0 & 121.1 & 21.0 & 2.7 \\
$\epsilon$ Car 	& $2.75\times10^{-16}$ & $1.93\times10^6$ & B2:V+K3:III & 274.29 & -12.60 & 185.5 & 270.1 & & \\
$\epsilon$ Lup	& $2.84\times10^{-16}$ & $1.91\times10^6$ & B2IV-V & 329.23 & 10.32 & 157.0 & 191.3 & & \\
$\zeta$ Tau	& $5.04\times10^{-16}$ & $1.91\times10^6$ & B1IVe\_shell & 185.69 & -5.64 & 136.4 & 174.0 & & \\
$\delta$ Cen	& $2.98\times10^{-16}$ & $1.87\times10^6$ & B2Vne & 296.00 & 11.57 & 174.9 & 209.6 & & \\
$\delta$ Cru	& $4.58\times10^{-16}$ & $1.82\times10^6$ & B2IV & 298.23 & 3.79 & 139.5 & 207.9 & $<20.2$ & $<0.4$ \\
$\nu$ Cen 	& $5.56\times10^{-16}$ & $1.75\times10^6$ & B2IV & 314.41 & 19.89 & 124.3 & 190.1 & & \\
$\beta$ CMa	& $3.49\times10^{-16}$ & $1.62\times10^6$ & B1II-III & 226.06 & -14.27 & 151.1 & 191.7 & 18.26 & 0.0039 \\
$\theta$ Car 	& $3.91\times10^{-16}$ & $1.57\times10^6$ & B0Vp & 289.60 & -4.90 & 139.7 & 205.4 & 20.26 & 0.42 \\
$\gamma$ Peg	& $3.55\times10^{-16}$ & $1.50\times10^6$ & B2IV & 109.43 & -46.68 & 143.9 & 264.7 & $<20.2$ & $<0.4$\\
$\epsilon$ CMa	& $4.73\times10^{-16}$ & $1.49\times10^6$ & B1.5II & 239.83 & -11.33 & 124.2 & 291.9 & 17.78 & 0.0016\\
BD+75 325 	& $3.44\times10^{-16}$ & $1.49\times10^6$ & sdO5 & 139.51 & 31.25 & 145.4 & 277.0 & & \\
$\epsilon$ Cen	& $4.03\times10^{-16}$ & $1.41\times10^6$ & B1III & 310.19 & 8.72 & 131.1 & 205.8 & $<19.9$ & $<0.2$ \\
$\zeta$ Cas	& $2.87\times10^{-16}$ & $1.32\times10^6$ & B2IV & 120.78 & -8.91 & 150.0 & 170.1 & & \\
Feige 24 	& $5.98\times10^{-16}$ & $1.15\times10^6$ & DA1+dM C & 165.97 & -50.27 & 77.7 & 191.3 & 18.45 & 0.012 \\
G191B2B 	& $1.12\times10^{-15}$ & $9.73\times10^5$ & DA.8 C & 155.95 & 7.10 & 52.5 & 143.90 & 18.18 & 0.0094 \\
HZ43 		& $2.23\times10^{-16}$ & $2.49\times10^5$ & DAwk D & 54.11 & 84.16 & 60.3 & 453.5 & 17.93 & 0.0046 \\
\\
 {\rm Total:}   & $5.69\times10^{-13}$ \\
\enddata
\noindent $^a$ Unless otherwise noted, the stellar properties listed here are taken 
from SIMBAD (https://simbad.u-strasbg.fr/simbad/sim-fid). \\
\noindent $^b$ Estimated \HI\ photoionization rate outside the LIC.
Note that the EUV spectra of these stars are based on model fits to their FUV spectra and are likely underestimates
(cf., for $\epsilon$~CMa the estimated EUV flux derived by Shull et~al. (2025) is $\sim 30 \times$ larger than given here). \\
\noindent $^c$ Estimated \HI\ photoionization flux at 1~pc from the star; the table is ordered by this value.
Note that the EUV spectra of these stars are based on model fits to their FUV spectra and are likely underestimates. \\
\noindent $^d$ Distance to the outer wall of the LHB in the star direction, as determined by O'Neill et~al. (2024).  \\
\noindent $^e$ Interstellar \HI\ column density (where available) from Diplas \& Savage (1994), Shull et~al. (2025),
Cassinelli et~al. (1996), Lemoine et~al. (2002), Kruk et~al. (2002), and Dupuis et~al. (1995), and \HI\ average number density
from dividing the column density by the distance to the star. \\
\end{deluxetable*}

\section{Acknowledgments}
We are grateful to D.~Hall for kindly providing the radiative transfer code used to calculate the brightness of backscattered solar Ly$\alpha$ and to A.~Goodman for helpful suggestions.
We thank the New Horizons team for support; in particular, Gabe Rogers, Debi Rose, and Jon Pineau for planning and executing
of the all-sky Ly$\alpha$ observations.

This research was supported by NASA New Horizons contract NASW02008 and by SwRI Internal Research Project R6411.

\clearpage
\bibliographystyle{aasjournal}

\end{document}